\documentstyle[12pt]{article}
\makeatletter
\@addtoreset{equation}{section}
\makeatother

\begin{document}
\thispagestyle{empty}
\begin{flushleft}
\begin{small}
LMU-TP 7/89 \\
RBI-TP 4/89 \\
Final version: July 1992 \\
\end{small}
\end{flushleft}
\vspace{1.5cm}
\begin{center}
\begin{Large}
\begin{bf}
Selected Topics on Rare Kaon Processes \\ {\large -in the Standard Model
and Supergravity-}\\
\end{bf}
\end{Large}
\vspace{1.2cm}
\begin{Large}
Neven Bili\'c\footnote{On leave of absence
from the Rudjer Bo\v{s}kovi\'c
Institute, Zagreb, Croatia.
Address after September 1, 1992: Department of Physics,
University of Cape Town, Private Bag Rondebosch 7700,
South Africa. E-mail: bilic@physci.uct.ac.za .}\\
{\small
Fakult\"at f\"ur Physik, Universit\"at Bielefeld,\\
Postfach 8640, D-4800 Bielefeld 1, F. R. Germany\\}
\vspace{.4in}
Branko Guberina\footnote{E-mail: guberina@thphys.irb.hr}\\
{\small Theoretical Physics Department, Rudjer Bo\v skovi\'c
Institute,\\
 Zagreb, Croatia\\}
\end{Large}
\begin{large}
\vspace{1.5cm}
To be published in {\bf Fortschr. der Physik}
\end{large}
\end{center}
\vspace{1.5cm}
\begin{center}
{\bf Abstract}
\end{center}
\begin{quotation}
\noindent
Rare kaon processes appear to be particularly suitable to study the extensions
of the standard model, especially if the possibility for eventual
direct evidence becomes unlikely. In this review,
we discuss  processes
that are important as a test of either
the standard model or supergravity.
Moreover, some of these  are important even for
both the standard model and for supergravity.

Particular attention is paid to the reduction of uncertainties
in  the calculation, especially
the ones coming from the confinement effects. Recent approaches,
such as chiral perturbation theory, the large $N_c$-expansion,
QCD sum rules and lattice QCD, are discussed.
This is found to be the best strategy in view of the
fact that supersymmetric effects are rather
tiny.

\end{quotation}
\newpage
\tableofcontents

\newpage
\section
{ Standard Model and Supersymmetry}

\subsection
{Introduction}

 Three basic interactions (electromagnetic,
weak, and strong) seem to be reasonably (even very
well) described by a theory known  as {\it the standard
model}\cite{b01}-\cite{b10}.
The fourth known interaction, gravitation, is not
included. Its classical version, the general theory
of relativity, describes macroscopic phenomena rather
well; the quantized theory of gravity, in spite of many attempts,
still does not exist as a self-consistent theory.
Fortunately, gravitational interactions at energies of
present experiments do not play any important role.
Therefore, the standard model may be tested to a high
degree of accuracy.

Up to now there has been no confirmed disagreement
between the standard model and experiment.
The standard model describes many phenomena very well;
actually, the description is too good for the taste
of many theoretical physicists. This `unhappiness'
arises from the answer to our question about the standard
model: the standard model is not a satisfactory
theory for the reasons to be discussed later.

Theoretically, one tries to construct theories which go
beyond the standard model and improve the drawbacks of the
standard model in some aspects. This includes a desirable reducing of the
number of parameters, improvement in convergence properties
and, particularly, the understanding of the Higgs sector
which, in itself, at least in the minimal version of the standard
model, shows unpleasant behavior at the quantum level.
Clearly, theories which go beyond the standard model\cite{b11},
bring new physics into play - usually a plethora of new
particles. Naively, this looks very promising because
new particles may, at least in principle, be detected
in one or another way. However, almost perfect agreement
between the standard model and experiment requires
that new physics should have very little impact
on low-energy phenomenology.
This  means that new
particles are rather heavy. Therefore, their direct production
and detection at large accelerators might finally appear
very difficult, especially if the present upper limits on masses
of new particles become considerably larger.

An alternative but complementary way of searching for
new physics are low-energy experiments with a high
degree of accuracy. This way is based on the following points.

(i) There are processes that appear at higher order in
electroweak interactions and which are very sensitive to
`impurities' caused by new physics. A special role is
here played by rare kaon processes. These include
rare kaon decays, CP violation in the kaon system,
$\epsilon '/\epsilon$ parameter, etc.

(ii) Significant progress has been made in reducing the
uncertainties in theoretical predictions in the standard
model. These uncertainties are usually connected with the
treating of hadrons, i.e., with the problem
of QCD confinement. A few new, rather sophisticated
approaches have  been addressed to this problem:
QCD lattice calculations\cite{b12,b13}, the large-$N_c$
approach\cite{b14}, and the QCD-duality approach\cite{b15}.

Since the effects of new physics appear to be rather tiny,
the importance of reducing the uncertainties in
calculations and/or errors in experiments is
{\it conditio sine qua non} for the future progress.

A number of experiments are presently running
at BNL, KEK, CERN and FNAL. Kaon beams are usually produced
from high-energy proton beams colliding on fixed targets.
Exception is the CP-LEAR project at CERN where low-energy $p\bar p$
collisions are used to study kaons in  the final state,
especially CP violation and CPT tests.

A clean, intense source of K-mesons is expected in $\phi$-factories,
like DA$\Phi$NE in Frascati, which is to  be built in the next
few years\cite{b16}.

A few proposals for K-factories have been considered in the
last few years, like European Hadron Facility\cite{b17} (abbandoned)
and TRIUMPH\cite{b18} with extremely
intense beams (planned after 1995).

{}~\\

{\it Rare Kaon Decays}
{}~\\

\addcontentsline{toc}{subsubsection}{Rare Kaon Decays}
The kaon system has proved to be the graveyard of many
wrong theories. The history goes from the
$\tau - \theta$ puzzle, CP violation, the
$\Delta I=1/2$ rule, the absence of $\Delta S=0$
neutral currents to the present search for lepton-flavor
violation and the existence of the fourth
generation. Last but not least, new ideas,
e.g., composite models and/or supersymmetry have to pass hard
tests in kaon physics.

The standard model (SM), based on the gauged
$SU(3)\times SU(2)\times U(1)$ theory with the
minimal Higgs sector, can be proved/disproved in the following ways:

(i) Going to higher energies and looking whether the SM
still works; this is basically the approach of
collider physics.

(ii) Precision measurements: measuring the parameters
of the SM highly accurate and looking for
discrepancies.

(iii) Searching for processes which are
{\it suppressed} or {\it forbidden}
in the standard model. This basically defines
the physics of {\it rare processes}, which are
the subject of this paper.

An example of forbidden process is
\begin{equation}
K_L\rightarrow \mu\; e,
\end{equation}
and, clearly, its observation at any level
would be a clean signal for new physics.
The present experimental limit is\cite{b19}
\begin{equation}
BR (K_L\rightarrow \mu\; e) < 0.94\times 10^{-10}.
\end{equation}
Other examples of $K$ decays that are forbidden in the SM,
but are {\it predicted} in new models are
\begin{eqnarray}
K^+\rightarrow \pi^+\mu^+ e^-, \nonumber \\
K_L\rightarrow \pi^0\mu e, \nonumber \\
K^+\rightarrow \pi^+ X^0,
\end{eqnarray}
where $X^0$ is a light scalar or pseudoscalar.

There are many processes in kaon physics that are
strongly suppressed in the standard model,
but are not totally forbidden. We list typical
examples.

The first process, $K_L\rightarrow e^+e^-$, is especially
sensitive to the Higgs sector of new physics since,
unlike in the SM, there would be no large
helicity suppression.

The $K_L\rightarrow \pi^0 e^+e^-$
process is suppressed by  a factor $\sim 10^{-6}$ owing
to the GIM mechanism and, in addition,
by a factor of $\sim 10^{-5}$ owing to CP conservation.

The $K^+ \rightarrow \pi^+\nu\bar\nu$ is probably
the cleanest test of higher-order electroweak
interactions {\it in} the standard model.
This particular process belongs to the class of
more general $K$ decays of the type
\begin{equation}
K^+\rightarrow \pi^+ + \;nothing,
\end{equation}
where `nothing' denotes any light neutral particle(s)
that cannot be detected.

Particularly interesting is the process
\begin{equation}
K^+\rightarrow \pi^+ + ? + ?'.
\end{equation}
This channel receives contributions from standard
decays (i.e., $K^+\rightarrow \pi^+ \nu\bar\nu$ )
and from the new physics beyond the standard model.
Being practically a short-distance phenomenon,
it has the advantage with respect to processes
such as $K_L\rightarrow \mu\bar\mu$ or
$K^+\rightarrow \pi^+ e^+ e^-$, where long-distance
contributions enter the game. Also, any
mechanism for lepton flavor violation
(horizontal gauge bosons, leptoquarks, etc.) would
lead to $K^+\rightarrow \pi^+ \nu_e \bar\nu_\mu$.
In addition, flavor-changing neutral currents in N=1
supergravity would give a contribution
to $K_L\rightarrow \pi^+ \nu\bar\nu$
{\it in addition} to the SM contribution.

Finally, the direct evidence for superparticles would be
possible in processes such as
\begin{equation}
K^+\rightarrow \pi^+ \tilde{\gamma}\tilde{\gamma},
\end{equation}
provided the photinos $\tilde{\gamma}$ are light enough
to be produced. Last but not least, the existence
of the fourth generation would certainly
influence this rare decay.

{}~\\

\subsection
{Limits of the Standard Model and Supersymmetry}

The standard model is a theory with a lot of parameters
which are not explained by the model itself. These include
lepton and quark masses, Yukawa couplings, mixing angles,
etc., whose pattern is not understood.
 Why have left-handed fermions so far appeared in $SU(2)$
doublets and right-handed ones in $SU(2)$ singlets?
 Why are there three colors, and
why is the electric charge quantized?
 The family problem is not understood: why three
generations of quarks and leptons?, etc.

Searching for answers one is forced to go beyond the standard
model. Among different theories, {\it supersymmetry} is
particularly interesting\cite{b20}-\cite{b47}; it is the symmetry of bosons
and fermions, whose theoretical motivation
we discuss in the following.

(i) Experience teaches us that nature obviously prefers
gauge theories and supersymmetry is the next logical gauge theory.

(ii) The spin degree of freedom is naturally contained only
in supersymmetry.

(iii) Supersymmetric theories are much better behaved mathematically,
and are even finite in some cases.

(iv) Locally gauged supersymmetry relates the SUSY
generators to the generators of space-time transformations leading
to {\it supergravity}, i.e., to a natural coupling of SUSY to
gravity.

The standard model can hardly be considered as a fundamental theory
since it contains, e. g., QED, which is not an asymptotically
free theory. Its interactions must become strong
at some higher energy scale. This is considered as a hint that one
has to treat the standard model rather as a low-energy
effective theory of a more fundamental one.

{}~\\

{\it The Higgs sector}
{}~\\
\addcontentsline{toc}{subsubsection}{The Higgs sector}

The Higgs sector gives additional hints which cast doubt upon
the fundamental character of the standard model.
The minimal Higgs sector needed to construct the standard
model contains one complex Higgs doublet $\Phi$ with the
potential
\begin{equation}
V(\Phi )=\lambda [\Phi^{\dagger} \Phi - \frac{1}{2\sqrt 2 G_F} ]^2 .
\end{equation}
After spontaneous symmetry breaking, this leads
to one physical Higgs field, with the mass
\begin{equation}
M_H^2 = \frac{\sqrt 2}{G_F} \lambda,
\end{equation}
with an unknown coupling constant $\lambda$.

As  far as the Higgs mass is concerned, it cannot be taken arbitrarily
large.   Radiative corrections
lead to a lower bound:

\begin{equation}
M_H \geq 6.6\; GeV.
\end{equation}
The various analyses \cite{b01,b04} show that the upper
limit on the Higgs mass is
\begin{equation}
M_H \leq 1\; TeV.
\end{equation}
Once the upper limit, $M_H \leq $ a few $TeV$, has been set,
the huge radiative corrections to
the mass of the Higgs boson provide an additional
hint that one needs a new physics
beyond the standard model. These corrections arise,
e.g., from different loops, which turn out to be
quadratically divergent, as is typical of scalar theory,
i.e., $\delta M_H^2 \sim \Lambda^2$.
Since it would be unnatural to expect
the corrections to be larger than the upper limit on
the Higgs mass, one expects the new physics
to give an {\it effective cutoff} scale
below a few $TeV$.

The idea is to use a higher symmetry to eliminate
quadratic divergences in the Higgs mass
corrections. Here supersymmetry appears to be {\it the}
proper theory because it can in principle eliminate
the problem of quadratic divergences.
The mechanism is rather simple:
In supersymmetry the loops of normal particles
are always accompanied by the loops of superpartners.
The extra minus sign, appearing because
of the fermion loop, leads basically
to the cancellation of divergences. This nice
property persists as long as the imposed supersymmetry
is exact. Since supersymmetry necessarily has to be
broken, the requirement of approximate
cancellation of divergences imposes a constraint
on the masses of superparticles.

The Higgs mechanism brings in the vacuum expectation value
$v$ which determines $M_W$ and fermion masses:
it is $v \simeq 250 \;GeV$. This value
is not derived in the standard model itself and needs
a fundamental explanation. Any attempt made
so far to explain $v$ in terms of higher symmetries
brings us to the {\it fine-tuning} and the
{\it hierarchy} problem.

Let us call $\mu_{weak}$ a scale at which $SU(2)\times U(1)$
breaking takes place, and $M$ a scale where a
fundamental theory becomes relevant.
Typically, this is a very high scale, e.g.,
$M_{GUT} \sim 10^{14} \; GeV$, $M_{Planck} \sim 10^{19}\;GeV$,
etc. Now, in the fundamental theory, the mass of the
Higgs particle is evaluated at the fundamental
scale $M$, and $M_H^2(M)$ has to be scaled
down to the weak scale $\mu_{weak}$.
Typically, the expression is
\begin{equation}\label{17}
M_H^2 (\mu_{weak}) =
M_H^2 (M) + const. \;\; g^2 \int_{\mu_{weak}}^{M} dp^2
+ \Re (M)g^2 + {\cal O} (g^4),
\end{equation}
where $\Re (M)$ grows almost as $\ln M$ when $M\rightarrow \infty$,
and $g$ is a coupling constant.
The second term in (\ref{17})  diverges quadratically
as $M\rightarrow \infty$.

In order to have $M_H^2 (\mu_{weak})\ll M^2$, one has
to fine-tune the parameter $M_H^2(M)$ extremely
accurately to cancel the second term in (\ref{17})
which is if order $M^2$.
This is known as the {\it fine-tuning problem} or
the {\it naturalness problem}.
Suppose, for example, that the standard model is
valid up to a GUT scale of $10^{15}\;GeV$ or even
up to the Planck scale of $10^{19}\;GeV$. If
these two scales plus the weak  scale were input
into the theory, the Higgs mass would have
to be chosen with  an accuracy of $10^{-34}$
compared with $M_{Planck}$.
Clearly, the natural value for $M_H^2(\mu_{weak})$ is
$\sim {\cal O} (M^2)$. Therefore, the fundamental question
is actually the {\it hierarchy problem}:
why is
$\mu_{weak} \ll M^2$\cite{b05,b07,b12}? Even if the question of fine-tuning
were solved in a satisfactory way, the hierarchy
problem  would still have to be solved.

As we have already mentioned, one has to impose constraints on
possible supersymmetry breaking to preserve
the cancellation of divergences. Such a symmetry breaking
is called {\it soft}. An example of soft
symmetry breakdown is a spontaneous symmetry breaking.
Clearly, with softly broken SUSY, the cancellation among diagrams
is partial. The finite results obtained are related
to the SUSY breaking scale $M_{susy}$.

A naive expectation would be that $M_{susy}$
should be of the order of $\mu_{weak}$. This really has
to be the case if SUSY is broken explicitly.
However, if the supersymmetry breakdown is soft,
then the splittings among supermultiplets are
proportional to $gM_{susy}$, $g$ being the
coupling constant. This happens in supergravity:
a soft breakdown may appear at $M_{susy}\sim 10^{11}\;GeV$,
and mass splittings and $M_W$ could still be
of the order of $100\;GeV$.

However, the `small' scale, $M_{susy}\sim 10^{11}\;GeV$,
is still very large so that gravity cannot be neglected.
Clearly, at such a scale all particles would have
a gravitational interaction and the mass splitting would
be of the order
\begin{equation}
\frac{M_{susy}^2}{M_{Planck}} \sim 10^3\;GeV,
\end{equation}
i.e., comparable with the scale of electroweak symmetry
breaking. The inverse Planck mass is basically the
gravitational constant, $\kappa \sim M_{Planck}^{-1}$.
This leads naturally to local supersymmetry or
supergravity.

In addition to {\it graviton} (spin 2) there also appears
{\it gravitino}, its superpartner (spin 3/2), with the mass
of the order
\begin{equation}
m_{3/2}\sim \frac{M_{susy}^2}{M_{Planck}},
\end{equation}
and the breakdown of supergravity induces the electroweak
breakdown with $M_W$ typically of the order of
$m_{3/2}$.

{}~\\
\newpage
\section
{Low-Energy $N=1$ Supergravity}

\subsection
{Soft Supersymmetry Breaking}

After the spontaneous breaking of a local $N=1$ supergravity one is
left with a low-energy theory which is an explicitly broken global
supersymmetry plus some `soft-breaking' terms. the theory is the
supersymmetrized minimally extended standard model.

The full lagrangian of the SM
 contains superpartners as well as the standard
particles. In the supersymmetric limit, the minimal
model of supersymmetry has the gauge group
$SU(3)\times SU(2)\times U(1)$, with three
generations of left chiral matter superfields
\begin{equation}
{\bf Q}(3,2,1/3),\;\;{\bf U}^c (\bar 3,1,-4/3),\;\;
{\bf D}^c (\bar 3,1,2/3),\;\;{\bf L}(1,2,-1)\;\;
{\bf E}^c (1,1,2)
\end{equation}
and two Higgs superfields
\begin{equation}
{\bf H}(1,2,1)\;\;\;and\;\;\;{\bf H}' (1,2,-1).
\end{equation}
The superpotential contains the most general
gauge-invariant couplings which preserve matter parity
and contains only the fields required by supersymmetrization:
\begin{equation}
W = h_U^{ij} {\bf H}{\bf U}_i^c{\bf Q}_j +
h_D^{ij} {\bf H}' {\bf D}_i^c {\bf Q}_j +
h_E^{ij} {\bf H}' {\bf E}_i^c {\bf L}_j +
\mu {\bf H} {\bf H}'.
\end{equation}
The couplings $h_U$, $h_D$ and $h_E$
are $3\times 3$ matrices in the generation space (i,j=1,2,3).

The supersymmetric lagrangian is built in such a way that
certain symmetries and conservation laws are
respected. For example, the baryon number $B$ and the lepton
number $L$ are conserved. This automatically excludes
terms such as
$({\bf H}{\bf L})_{\theta\theta},
({\bf U}^c {\bf D}^c {\bf D}^c )_{\theta\theta},
({\bf L}{\bf L}{\bf E}^c )_{\theta\theta}$, etc.

The important new discrete symmetry introduced is $R$-parity,
defined as
\begin{equation}
R = (-)^{3(B-L)} (-)^F ,
\end{equation}
where $F$ is the fermion number.
This symmetry is introduced to  enforce baryon and
lepton number conservation. With R-parity
conserved, normal particles are R even, while all
superparticles are R odd. As a consequence, the lightest
superparticle (LSP) is stable.

A {\it global R invariance} is achieved in the limit
$\tilde m_i \rightarrow 0$ and
$h_U =h_D =h_E = 0$, i.e.,
a chiral symmetry protects gauginos from getting mass in all
orders in perturbation theory.

The term $\mu {\bf H}{\bf H}'$ is introduced in order to avoid
an unwanted massless axion. In the limit
$\mu \rightarrow 0$ there exists an exact Peccei-Quinn
symmetry, $U(1)_{PQ}$, which gets spontaneously broken when
the Higgs fields develop vacuum expectation values.
{}~\\

As we discussed earlier, the most convenient way of breaking supersymmetry
softly is to couple it to a hidden sector of $N=1$ supergravity.
The situation is much simplified by the assumption of
having a flat K\"ahler metric.
There are a few types of soft SUSY breaking.

(i) A cubic gauge-invariant polynomial in complex scalar fields
(trilinear scalar couplings):
\begin{equation}\label{33}
[ \xi_U {\bf H}{\bf U}^c {\bf Q} +
\xi_D {\bf H}' {\bf D}^c {\bf Q} +
\xi_E {\bf H}' {\bf E}^c {\bf L} +
\mu B {\bf H} {\bf H}' ]_A + h.c.,
\end{equation}
where $\xi$ are $3\times 3$ flavor matrices.
This part of soft breaking contributes to squark and slepton
mass matrices.

(ii) Gaugino Majorana  terms:
\begin{equation}
\frac{1}{2} \sum_\alpha \tilde m_\alpha \lambda_\alpha
\lambda_\alpha + h.c.
\end{equation}
These terms obviously give masses to all gauginos, i.e., to
wino, zino, and gluino.

(iii) Mass terms of the scalar fields $z_i$ of the chiral
superfields:
\begin{equation}
M_{ij}^2 z_i^* z_j + h.c.
\end{equation}
These terms give masses to all scalar superparticles, i.e.,
squarks, sleptons, and higgs.

(iv) The last term
\begin{equation}
\mu B {\bf H} {\bf H}'
\end{equation}
enters the Higgs potential.

In addition, if one assumes that the low-energy
theory is a Grand Unified Theory, it would lead to a set
of relations among the soft couplings and masses.
Under the assumption that a grand unification appears
at some scale $M_X$, a natural assumption is the universal
gauge coupling. All soft-breaking terms are assumed
to be parametrized at $M_X$ by a universal gaugino mass $m_{1/2}$,
all scalar masses are given by $m_0$ at $M_X$, and
there is a universal trilinear scalar coupling $A$.

The above scalar and gaugino mass terms are allowed
since gauge invariance is preserved. However, gauge invariance
forbids mass terms for quarks, leptons, and gauge bosons - they
will acquire their masses through spontaneous gauge symmetry
breaking. Incidentally, only such particles have
been seen until now.

The only peculiar behavior as far as the origin of mass
is concerned shows higgsino. While the higgs particle acquires
mass from the usual weak spontaneous breakdown, the higgsino gets mass
both from the $\mu$-term in ${\cal L}_{susy}$ and from the weak
$SU(2)\times U(1)$ symmetry breaking.

\subsection
{Flavor-Changing Neutral  Currents}

The existence of two scalar doublets in the supersymmetric
extension of the standard model does not lead
to flavor-changing neutral currents since the
underlying supersymmetry forbids it  at the
tree level.

However, one-loop radiative corrections induced by the
charged scalar would generally lead to the couplings between
squarks, quarks, and gluinos containing
flavor-changing parts. In addition, the hidden
sector of supergravity, needed for the super-Higgs
mechanism will bring new phase into the theory.
If a typical SUSY breaking scale were of the order
of $10^{10}\;GeV$,
one might expect large renormalization effects.

{}~\\

{\it Squark Mass Matrix}
{}~\\
\addcontentsline{toc}{subsubsection}{Squark  Mass Matrix} \\

After $SU(2)\times U(1)$ symmetry is spontaneously
broken, the potential invariant under {\it global}
SUSY brings into play squark mass terms
$M_Q^{\dagger} M_Q$, $M_Q$ being the squark mass matrix and
with mass terms in ${\cal L}$ typically of
the form
\begin{equation}
\tilde q _L^{\dagger} \tilde q_L + (L\rightarrow R),
\end{equation}
$\tilde q_{L,R}$ being the squarks (L- and
R- handed, respectively). In the minimal
N=1 supergravity extension of the SM,
the soft-breaking terms contribute in such a way
that squark masses are degenerate for
different generations
\begin{equation}
\mu_L^2 \tilde q_L^{\dagger} \tilde q_L +
\mu_R^2 \tilde q_R^{\dagger} \tilde q_R,
\end{equation}
and $\mu_{L,R}^2$ are of the order of
the global SUSY breaking scale,
$\mu\sim 10^2\;GeV$, and usually
taken to be equal to the gravitino
mass $m_{3/2}$.

Trilinear interactions between
the Higgs field and two squark fields (cf. Eq.(\ref{33}))
contribute with a strenght $A$ and are proportional
to $\mu$. The parameter $A$ may, in principle,
be calculable from the hidden sector of
supergravity.

Left-right squark mixings, i.e., the mass
terms of the type
 $\tilde q_L^{\dagger} \tilde q_R +\;h.c.$
are induced through the electroweak spontaneous
breakdown and are proportional to $M_Q$.

Finally, radiative corrections to the down-squark masses
are induced by the charged Higgs and the higgsino
loop and are proportional to
$M_u^{\dagger} M_u$, $M_u$ being the up-quark mass
matrix.

The down-squark mass term becomes
(taking into account  quantum corrections)
\begin{equation}
M_D^2 \simeq
\left(
\begin{array}{cc}
\mu_L^2 {\bf 1} + M_d M_d^{\dagger} + cM_u M_u^{\dagger} &
A^{*} m_{3/2} M_d \\
A m_{3/2} M_d^{\dagger} &
\mu_R^2 {\bf 1} + M_d^{\dagger} M_d
\end{array}
\right) .
\end{equation}
The parameter $c$ is negative and is usually assumed
to be ${\cal O} (1)$.
The parameter $A$ is generally a complex parameter;
it can be written as
\begin{equation}
A = |A| e^{-2i\phi_A} .
\end{equation}
The matrices $M_D^2$ and $M_d$ cannot be diagonalized
simultaneously by the same transformation.
Let us introduce 2 unitary matrices,
$\tilde U_1$ and $\tilde U_2$, such that
$\tilde U = \tilde U_1 \tilde U_2$
and
\begin{equation}
\tilde U_1 =
\left(
\begin{array}{cc}
e^{i\phi_A} U_L^d  &  0 \\
0  &  e^{-i\phi_A} U_R^d
\end{array}
\right) ,
\end{equation}
where $U_L^d , U_R^d$ diagonalize
the down-quark matrix. Then the matrix
$\tilde U$ diagonalizes the down-squark matrix:
\begin{equation}\label{44}
\tilde U^{\dagger} M_D^2 \tilde U = \tilde U_2^{\dagger}
\left(
\begin{array}{cc}
\mu_L^2 {\bf 1} + \hat{M}_d^2 +  cV^{\dagger}\hat{M}_u^2 V &
|A| m_{3/2} \hat{M}_d \\
|A| m_{3/2} \hat{M}_d &
\mu_R^2 {\bf 1} + \hat{M}_d^2
\end{array}
\right) \tilde U_2 .
\end{equation}
 $\hat{M}_{u,d}$ are diagonalized
$M_{u,d}$ matrices and
\begin{equation}
V = U_L^{u\dagger} U_L^d
\end{equation}
is the usual Kobayashi-Maskawa matrix.

If one neglects the left-right squark mixing,
 and
since $\hat{M}_u^2 \gg \hat{M}_d^2$,
a reasonable approximation for $\tilde U_2$ is
\begin{equation}
\tilde U_2 \simeq
\left(
\begin{array}{cc}
V^{\dagger}  &  0 \\
0  &  {\bf 1}
\end{array}
\right) ,
\end{equation}
which leads to
\begin{equation}\label{47}
\hat{M}_D^2 \simeq
\left(
\begin{array}{cc}
\mu^2 {\bf 1} + c\hat{M}_u^2  &
|A| m_{3/2} V \hat{M}_d^{\dagger} \\
|A| m_{3/2} \hat{M}_d V^{\dagger}  &
\mu^2 {\bf 1} + \hat{M}_d^2
\end{array}
\right) ,
\end{equation}
The current- and mass- eigenstates are connected via
\begin{equation}
\tilde d = \tilde U
\left(
\begin{array}{c}
\tilde d_L^{(0)} \\
\tilde d_R^{(0)}
\end{array}
\right)
\end{equation}
and the latter induce flavor-changing couplings.

\newpage
\section
{Recent Approaches in Calculating Hadronic
Transition Amplitudes}

\subsection
{Chiral Perturbation Theory - Basics}

Chiral perturbation theory (ChPT) represents a viable alternative low
energy theory of strong and electroweak
interactions\cite{b48}-\cite{b75}. Its importance
as an alternative approach to the usual formulation of the standard model
is especially evident in hadronic processes where the lack
of knowledge of QCD confinement is blurring our view of electroweak
interactions. This is even more pronounced in kaon physics,
since kaons and pions, being pseudo-Goldstone bosons,are even less
reliably described in phenomenological quark models.

This theory has recently become important in connection with some
other alternative approaches, such as QCD hadronic duality sum
rules\cite{b56}-\cite{b58}, QCD simulation on lattice\cite{b54}
and in the large-$N_c$ approach of Bardeen et al.\cite{b63}.
In all of these approaches, ChPT is used in
some way or another: in
duality sum rules, the tree level chiral realization of hadronic
currents and/or operators is used in the parametrization of the hadronic
side of the sum rule. In the large-$N_c$ approach of ref.\cite{b63}, the
chiral representation of currents is used to construct  weak transition
operators at zero momentum, which
is needed in order to describe the operator evolution at large distances.
Finally, the lattice calculation of weak matrix
elements is (presently) performed for off-shell
transitions, such as the  $K\rightarrow \pi $ transition. The relation
with physical amplitudes ($K\rightarrow \pi\pi$) is achieved using the tree
level ChPT relation between the above transitions.

Chiral perturbation theory is based on our knowledge of the fundamental
symmetries of the QCD lagrangian, e. g., the softly broken chiral
$SU(3)_L\times SU(3)_R$ symmetry. The formulation of ChPT is
based\cite{b51,b52} on the folowing
{\it Ansatz}:
in any given order in perturbation
theory,
the most general lagrangian consistent with
a given symmetry would result
in the most general $S$ matrix,
 consistent with incorporated symmetry and
 field
theory conditions on analiticity, perturbative unitarity, etc.
Formulated in this way, ChPT becomes a quantum field theory.

The strong lagrangian at order $p^2$ (with minimal number of
derivatives) is given by the nonlinear $\sigma$ -model
\begin{equation}
{\cal L}^{(2)} = \frac{f^2}{8}
tr(\partial_\mu U \partial^\mu U^{\dagger} )
= \frac{f^2}{8} g_{ab} (\phi ) \partial_\mu \phi^a
\partial^\mu \phi^b
\end{equation}
for a massless field, with the invariant metric
\begin{equation}
g_{ab} (\phi ) = tr (\partial_a U\partial_b U^{\dagger}).
\end{equation}
$U$ is the unitary matrix field
\begin{equation}
U(\phi ) = \exp i\frac{2}{f} \Phi .
\end{equation}
$\Phi = \frac{1}{\sqrt 2} \lambda^a \phi^a$ is
given by
\begin{equation}
\Phi =
\left(
\begin{array}{ccc}
\frac{\pi^0}{\sqrt 2} + \frac{\eta_8}{\sqrt 6} &
\pi^+  &  K^+  \\
\pi^-  &  -\frac{\pi^0}{\sqrt 2} + \frac{\eta_8}{\sqrt 6}
&  K^0 \\
K^-  &  \overline K^0  &  -\sqrt{\frac{2}{3}} \eta_8 .
\end{array}
\right)
\end{equation}

If the quark mass matrix ${\cal M}$ is different from zero,
all terms in ${\cal L}_{strong}$ would
pick up additional terms. Then, the strong lagrangian
 to the lowest order, reads
\begin{equation}\label{53}
{\cal L}^{(2)}_{strong} =
\frac{f^2}{8} tr (\partial_\mu U \partial^\mu U^{\dagger} )
+ v\;\; tr ({\cal M} U + U^{\dagger} {\cal M} ),
\end{equation}
with
\begin{equation}
\frac{4 v}{f^2} =
\frac{m^2_{\pi^+}}{m_u + m_d } =
\frac{m^2_{K^+}}{m_s + m_u }   =
\frac{m^2_{K^0}}{m_s + m_d }   = B_0 ,
\end{equation}
and $v$ is proportional to the quark
condensate, $v = -\frac{1}{4} <0 | \bar qq | 0>$.
The lowest order ${\cal L}^{(2)}_{strong}$ at the tree level
has basically two couplings, $f_\pi$ and $v$, which are
expected to be calculable in QCD.

The lagrangian ${\cal L}^{(2)}_{strong}$ of the nonlinear
$\sigma$-model is nonrenormalizable
and agrees with QCD only at the tree level (leading
behavior).\footnote{The choice
of the effective lagrangian is not unique. At the tree level,
both linear and nonlinear lagrangians lead to the proper
value of the two constants $f_\pi$ and $v$.
At the one loop level, however, they disagree.
The linear (renormalizable) $\sigma$ model gives
relations among the couplings which are in contradiction with experiment.
}
In order to get the full
strong lagrangian in ChPT
 one has to include
{\it all} possible terms in the lagrangian
and take account of all graphs in perturbation theory.
The classical theory is equivalent to
tree graphs of quantum field theory. However,
a consistent quantum field theory requires the
presence of loop graphs, since without loops, the theory
{\it violates unitarity}. For example, the
effective lagrangian ${\cal L}^{(2)}$
describes the scattering of pseudoscalar
mesons to order $p^2$ (tree graphs),
the parameters being the quark masses,
$f_\pi$ and $v$.
The unitarity demands that at  order $p^4$
the $T$-matrix involves cuts with
discontinuities, the contribution being
determined up to a polynomial in the
external momenta.
In field theory, these discontinuities
appear in the one-loop graphs from ${\cal L}^{(2)}$.

The inclusion of loops would generally lead
to infinities. To get rid of them, the theory should
allow for counterterms. Using the regularization
that preserves chiral symmetry (e.g., dimensional
regularization), one finds
that one needs counterterms of order $p^4$,
since loop-graphs coming from the $p^2$-lagrangian
are of that order. Furthermore, one finds
that all counterterms needed are contained
in ${\cal L}^{(4)}$. This is intuitively clear,
since we demand that regularization preserves
chiral symmetry. Using the terms
in ${\cal L}^{(4)}$ as counterterms
leads to finite results for all Green's
functions to one-loop order. However,
the renormalized graphs are unambiguous
only
up to a polynomial in the external momenta.

In order to include the electromagnetic interaction,
one has
to couple the quarks to hermitian external fields,
 as proposed by  Gasser and Leutwyler \cite{b52}.

The lagrangian ${\cal L}^{(2)}$
becomes
\begin{equation}
\stackrel{\circ}{{\cal L}
^{(2)}}  =
\frac{f_\pi^2}{8}
tr (D_\mu U D^\mu U^{\dagger} ) +
v\;tr( {\cal M} U + U^{\dagger} {\cal M} ),
\end{equation}
where $D_\mu$ is the covariant derivative
\begin{equation}
D_\mu =\partial_\mu U - ie {\cal A}_\mu [Q,U] ,
\end{equation}
and ${\cal A_\mu}$ is an external electromagnetic field.

\subsection
{Chiral Realization of Hadronic Weak Interactions}

The extension of ChPT to weak processes is straightforward.
However, the predictive power is often spoiled by
a large number of unknown coupling constants.

In weak $\Delta S=1$ kaon decays, CP violation, etc., one
has only a few channels and this makes it difficult to
go beyond the tree-level lagrangian.
Nevertheless, the chiral realization of weak
lagrangians\cite{b54} has been found to be very useful
as it `supports' other approaches, such as
 QCD lattice calculations,
QCD hadronic duality sum rules, and the large-$N_c$
expansion.

For example, to the order $p^2$, the octet
part of the $\Delta S=1$
weak lagrangian is given by\cite{b54}
\begin{equation}\label{61}
{\cal L}_{\Delta S=1}^{(2)} =
g_8 \tilde {\cal L}_8 + h_8 \Theta ,
\end{equation}
where $g_8$ and $h_8$
are unknown coupling constants, not
fixed by ChPT alone. The operators
$\tilde {\cal L}_8$ and $\Theta$ have the
following realization to the lowest
order in derivatives and masses
\begin{eqnarray}
\label{62}
\tilde {\cal L} =
\frac{G_F}{2\sqrt 2} s_1 c_1 c_3\;\;
tr (\Lambda \partial_\mu U \partial^\mu U^{\dagger} ) \nonumber \\
\Theta =
\frac{G_F}{2\sqrt 2} s_1 c_1 c_3\;\;
\frac{8v}{f^2}\;\;
tr (\Lambda U {\cal M} + \Lambda (U {\cal M})^{\dagger} ).
\end{eqnarray}
The operator $\Theta$ is the tadpole operator
which in general contributes  to the off-shell
Green's functions, but does not
contribute to the $S$ matrix.
For example, the operator $\Theta$
contributes to the amplitudes
${\cal A} (K\rightarrow \pi\pi)$,
${\cal A} (K\rightarrow \pi )$,
and
${\cal A} (K\rightarrow 0)$.
The first graph in Fig. 1-I gives
$\sim \frac{4}{3} \frac{m_s - m_d}{f^3} h_8$,
which is , however, exactly canceled
by the contribution of the second graph in Fig. 1-1.
On the other hand, if one writes the amplitude
${\cal A} (K\rightarrow \pi\pi)$
in terms of the amplitudes
${\cal A} (K\rightarrow \pi )$
and
${\cal A} (K\rightarrow 0)$
(the PCAC relation), one finds again that
the tadpole contributions in the last
two amplitudes cancel each other and
${\cal A} (K\rightarrow \pi\pi)$
is tadpole-free.

Recently, Shabalin\cite{b48} has suggested
an interesting and intriguing mechanism
that turns a quadratic GIM suppression
of the $d - s$ self-energy graph into a logarithmic
one, showing a clear enhancement of the $\Delta I=1/2$
transition. The effect  comes basically
from the leading logarithmic one-gluon
corrections to the bare graph.
However, it has been shown
by Guberina, Peccei, and Picek\cite{b49}
that the full QCD correction in the
leading logarithmic approximation (LLA)
reduces the tadpole contribution to a negligible
amount, which vanishes  in the chiral limit.

The physical amplitude
${\cal A} (K\rightarrow \pi\pi)$
is therefore proportional to $g_8$,
which can be extracted from experiment.
Numerically,
$| g_8 | \approx 5.1$ .
Next-to-leading corrections to $K\rightarrow 2\pi$ and
$K\rightarrow 3\pi$ have been calculated by Kambor {\it et al.}\cite{b65}.
The counterterms were fitted from the decay rates and slope parameters.
The lowest-order value of $g_8$ reduces by 30\%, whereas the
analogous $27$-plet constant remains practically unchanged.

{}~\\

{\it Effective Lagrangian for Radiative Kaon Decays}
{}~\\

\addcontentsline{toc}{subsubsection}{Effective
Lagrangian for Radiative Kaon
Decays}

The electromagnetic field may be introduced into
${\cal L}_{\Delta S=1}^{(2)}$ through a covariant
derivative
\begin{eqnarray}
\label{63}
{\cal L}_{\Delta S=1}^{(2)} =
\tilde g_8 tr\; (\Lambda D_\mu U
D^\mu U^{\dagger} ) , \nonumber \\
\tilde g_8 = \frac{G_F}{2\sqrt 2}
s_1 c_1 c_3 g_8 ,
\end{eqnarray}
In addition, one has to add to it
all possible terms of fourth order
in derivatives and/or external fields
allowed by chiral symmetry.
Finally one finds\cite{b53}
\begin{eqnarray}
\label{64}
{\cal L}_{\Delta S=1,\;em}^{(4)} =
-\frac{ie2\tilde g_8}{f_\pi^2}
F^{\mu\nu} \{ w_1\;tr(Q\Lambda \hat{L}_\mu \hat{L}_\nu )
+ w_2 \;tr (Q\hat{L}_\mu \Lambda \hat{L}_\nu ) \} \nonumber \\
+ \frac{e^2 f_\pi^2 \tilde g_8}{2}
w_4 F^{\mu\nu} F_{\mu\nu}\;
tr ( \Lambda Q U Q U^{\dagger} )
+ h.c. ,
\end{eqnarray}
and $\hat{L}_\mu$ is a left-handed current with
covariant derivatives.
 Only the terms relevant for radiative $K$
decays with one pion in the final state
are kept.
Again, $w_1$, $w_2$, and $w_4$ are
undetermined couplings.
To this order, one has to add
the contributing part of the strong + electromagnetic
lagrangian
\begin{equation}
\label{65}
{\cal L}_{strong+em}^{(4)}=
-ieL_9 F^{\mu\nu} tr\;
(QD_\mu UD_\nu U^{\dagger} +QD_\mu U^{\dagger} D_\nu U)
+ e^2L_{10} F^{\mu\nu}F_{\mu\nu} tr\;
(QUQU^{\dagger}) ,
\end{equation}
and the anomalous Wess-Zumino lagrangian
\begin{equation}
\label{66}
{\cal L}_{WZ}^{(4)}=
\frac{\alpha}{4\sqrt 2 \pi f^2} \epsilon_{\mu\nu\rho\sigma}
F^{\mu\nu} F^{\rho\sigma}
(\pi^0 + \frac{\eta}{\sqrt 3}) +
{\cal O} (\phi^3,\phi^5) ,
\end{equation}
linear in meson fields. This completes the tools
necessary to calculate  radiative $K$ decays, which
have been discussed in Chapter 4.

\subsection
{Large-$N_c$ Expansion in ChPT and Beyond}

{}~\\

{\it Large-$N_c$ Expansion in the Weak Sector in ChPT}
{}~\\

\addcontentsline{toc}{subsubsection}{Large-$N_c$ Expansion
in the Weak Sector in ChPT}

The large-$N_c$ limit can be taken consistently
for all currents and/or operators appearing in
the weak sector of ChPT. For example,
the coupling $g_8$ in (\ref{61}) is not determined in
ChPT. The value of $g_8$ is known only in the
large-$N_c$ limit, in which the weak lagrangian
(\ref{61})  is reduced to the product of bare currents.
With $g_8$ determined in the large-$N_c$ limit,
the amplitude ${\cal A} (K\rightarrow \pi\pi )$
coincides with  the large-$N_c$ vacuum saturation
approximation for the matrix elements of
local 4-quark operators.

 The $p^4$-level lagrangian
introduces many unknown coupling constants.
Typically, the amplitude receives contributions
from one loop graphs calculated with the $p^2$-lagrangian
and from tree graphs stemming from the $p^4$-lagrangian.
The unknown couplings in $p^4$-lagrangians are to
be used as counterterms needed to renormalize
possible divergences in loop integrals. The finite
part of counterterms depends on the renormalization
point $\mu$; this dependence cancels the
logarithmic $\mu$ dependence
of the renormalized loop-graph contribution.
If one uses  dimensional regularization, one
encounters only logarithmic divergences. Working instead
with cutoff regularization, one also encounters  quadratic
divergences. The latter, however, disapppear
after renormalization, as we shall discuss later.

The role of meson loops and/or counterterms is very important,
as can be seen by studying rare kaon decays (cf. Chapter 4).
The physical meaning of the counterterms (couplings) is the following:
they are remnants of the original quarks and gluons after
the latter are integrated out in
the functional integral. Therefore, they contain both
{\it short- and long- distance} physics, i.e.,
the effects of, e.g., vector mesons (long distance) and
hard gluons (short distance).
Of course, in some cases, such as the
$p^4$ couplings
$L_1,...,L_{10}$ in the strong+electromagnetic lagrangian,
they are perfectly  saturated by  low-lying meson
resonances. The situation is, however, quite
different for  weak couplings, e.g., the
constants $w_i$ in (\ref{65}). Once determined,
the renormalized couplings depend on the renormalization point $\mu$;
however, their  natural order of magnitude is $(4\pi )^{-2}$.

The large-$N_c$ expansion turns out to be a very
useful tool in treating many problems in strong
interactions. In the weak-interaction sector the situation
is somewhat different. The idea of the large-$N_c$
expansion\cite{b67}
is based on the expectation that the true expansion
parameter is not $1/N_c = 1/3$, but rather something
like $1/4\pi N_c$ or even $1/4\pi N_c^2$, as it happens in
QED, where an expansion in the coupling constant $e$
becomes an expansion in $\alpha = e^2 / 4\pi$.
Unfortunately, in the weak-interaction sector the
subleading $1/N_c$ corrections are often large,
even huge in some cases\footnote{The leading terms in the
large-$N_c$ expansion of the $K$-decay amplitudes
lead to the ratio
${\cal A} (K^0\rightarrow \pi+\pi^- )/
{\cal A} (K^+\rightarrow \pi^+\pi^0 )$
of the order 1, which is by an order of
magnitude smaller than the experimental result.}.

If one works with an explicit cutoff, the cutoff
dependence comes in  different ways\cite{b61}:
(i) A possible $\Lambda^4$ contribution is absent because
of chiral symmetry, (ii) quadratic cutoff dependence
is, by power counting and chiral symmetry, of the
form $\Lambda^2\times$ tree-level ${\cal L}^{(2)}$,
(iii) logarithmic terms are of the form
$\ln \Lambda^2\times$ tree-level ${\cal L}^{(4)}$.
It is then possible to absorb the cutoff dependence
in the redefinition of coupling constants. It has been
shown by Bijnens and Guberina\cite{b61} that the results obtained
coincide with the ones obtained using dimensional
regularization.

As an example, the $K^0 - \bar K^0$ mixing transition
amplitude appears to be\cite{b61}
\begin{equation}
< K^0 | {\cal L}_{s\bar ds\bar d} | \bar K^0 > =
(\frac{f^2}{2} + \frac{8g_1}{f^2})
m_K^2 - \frac{\Lambda^2}{16\pi^2}
5 m_K^2 + \cdots
\end{equation}
The leading term is of the order $N_c^2 \sim f^2$, i.e.,
it is the leading factorizable contribution. The new unknown
coupling $g_1$ is of the order $N_c$. The cutoff
dependence can be partially absorbed in the
definition of the renormalized coupling $f^{ren}$,
and the rest enters the renormalized coupling
$g_1^{ren}$. The last subtraction is consistent
with the large-$N_c$ behavior,
since $g_1$ and the quadratic divergence
are of the same order in $N_c$.
The logarithmic divergences are handled in the same way.

At the end, one can remove all quadratic and logarithmic
divergences in the renormalization procedure. The
logarithmic $\mu$ dependence is cancelled
by the corresponding $\mu$ dependence of the counterterms.
Both, the cutoff regularization and the dimensional
regularization
lead to the same result\cite{b61}.

 In the next subsection, we discuss the
Bardeen-Buras-G\'erard approach to the large-$N_c$ expansion,
and QCD hadronic duality sum rules.

\newpage

{\it Bardeen-Buras-G\'erard Approach versus ChPT}
{}~\\

\addcontentsline{toc}{subsubsection}{Bardeen-Buras-G\'erard
Approach versus ChPT}

The Bardeen-Buras-G\'erard (BBG) approach\cite{b63} is based on the large-$N_c$
expansion
of QCD (with quarks and gluons), which implies
the existence of an equivalent dual-meson representation.
The effective $1/N_c$ expansion looks in the latter
representation like a string theory.
The leading-$N_c$ theory contains infinite trajectories
of stable meson resonances.
Since it is expected that the low-energy theory
is only sensitive to low-lying states, one may
use a proper truncation of the full theory.
The proposal of Bardeen et al. is to use
a nonlinear $\sigma$ model, including loop effects, as
a first approximation. The strong lagrangian (\ref{53}) in ChPT,
discussed in preceding sections, now appears  as
the low-energy  truncation of QCD.

The correct infrared behavior of low-energy theory
is guaranteed by low-mass and low-momentum
loop contribution. The truncation is achieved
by introducing a physical momentum cutoff in loop
integrals.
As expected, the main contribution
comes from the quadratic cutoff $\Lambda^2$.
This is the crucial difference with respect to ChPT,
where the cutoff dependence is absorbed in the
renormalized quantities. Formally, the typical
next-to-leading order amplitude in the BBG approach
looks like
\begin{equation}
\label{71}
c^{-1}(\mu_{QCD}){\cal A} \sim
\Lambda^2 + b \ln \frac{m_K^2}{\Lambda^2} + \cdots ,
\end{equation}
which is to be compared with a typical amplitude
in ChPT
\begin{equation}
\label{72}
{\cal A} \sim w(\mu^2) + b \ln \frac{m_K^2}{\mu^2}
+ \cdots  .
\end{equation}
For $\Lambda \sim \mu$, the logarithmic terms are
the same, and the finite counterterms in (\ref{72}) are traded
for the quadratic cutoff in (\ref{71}). However,
this difference is only the formal one since the
amplitude (\ref{72}) is the full amplitude, whereas
 the BBG result (\ref{71}) is the
matrix element of the {\bf local operator}, i.e.,
the short-distance behavior is factorized out
in the form of the Wilson coefficient.
This makes the comparison between the BBG approach and ChPT
by no means trivial. In addition, the physical cutoff
in the BBG approach plays the role of a scale
that has to match the corresponding renormalization scale of QCD.
In the chiral limit, for example, one has to match
the quadratic $\Lambda^2$ behavior with the logarithmic
$\ln \mu_{QCD}$ one, which appears difficult\cite{b61}.
However, higher resonances are expected to improve matching,
smoothing the quadratic behavior of the scale $\Lambda$
and forcing the approximate
logarithmic behavior\footnote{
It is interesting to note\cite{b55} that one can
exactly control the $\mu$ dependence if one
calculates the leading $1/N_c$ behavior of the penguin
operator $Q_6 = -8\sum_q (\bar s_L q_R ) (\bar q_R d_L )$.
The leading term has the following chiral realization
\begin{equation}
Q_6^{ChPT} = -16 v^2 \frac{8L_5}{f_\pi^2}
tr\;(\Lambda \partial_\mu U \partial^\mu U^{\dagger} ).
\end{equation}
Now, $v^2$ is proportional to $(\overline m_q)^{-2}$,
where $\overline m_q$ is the running quark mass, whose QCD
behavior is $\overline m_q^2 \sim \hat{m}^2 \alpha_s
(\mu_{QCD}^2)^{9/11}$.
On the other hand, the leading $1/N_c$ behavior
of the Wilson coefficient is given by
$c \sim \alpha_s (\mu_{QCD}^2)^{9/11}$.
Obviously, doing a systematic $1/N_c$ expansion, one
achieves an exact cancellation of the
$\mu$ dependence.}.

The second difference with respect to ChPT, i.e.,  that no
counterterms were allowed in \cite{b63} means that, e.g., vector
mesons and higher resonances have to be added separately.
In ChPT, the latter are contained in counterterms.

As we mentioned, one would naively guess that the
quadratic cutoff would have to mimic counterterms
if both theories were equivalent. This, however, is not true for
the following reasons: (i) contrary to the cutoff-terms, the
counterterms contain long-distance effects as well as
short-distance ones. In the approach of \cite{b63}, the latter
are factorized in the form of Wilson coefficients;
(ii) the counterterms in ChPT also contain the effects of higher
resonances (vector mesons, etc.). As given in \cite{b63},
they have to be added separately.

In addition, ChPT would implicitly require the importance of counterterms
in order to keep the whole correction in weak amplitudes
moderate. Otherwise, the perturbative expansion would
break since
in some cases  loop corrections contain
 huge logarithms\cite{b68}.
This implies the importance of missing corrections due to vector
mesons, etc., in the BBG approach.

 From the above points it appears difficult to achieve one-to-one
correspondence between ChPT and the large-$N_c$ approach
of Bardeen et al. However, we would like to point out that
in spite of our inability to achieve it, this genuine approach
of Bardeen et al. leads to reasonable results. For example,
the result for the $B$ parameter, $B \sim 0.7$ is rather
stable as well as the result for the $K^0 \rightarrow
\pi^+\pi^-$ amplitude. The latter is actually  close
to the experimental value, i.e., the $\Delta I=1/2$ rule
seems to be explained in the BBG approach. The
$\Delta I=3/2$ amplitude, however, is sensitive to
the choice of the cutoff which induces a large uncertainty.

Recently, an exciting project has been started\cite{b66} in an
attempt to derive the low-energy chiral realization of the
$\Delta S=1,2$ operators by integrating out the quark fields
in a gluonic background. The spontaneous breaking of chiral
symmetry is triggered by  a source term, which is added to
the QCD lagrangian. The bosonization of the QCD lagrangians
generates the effective lagrangian of ChPT but with explicit
values for different couplings! Further work in this direction is
in progress.

\subsection
{QCD Hadronic Duality Sum Rules}

In the standard model, the effective hadronic weak lagrangian
 is usually given
in terms of local 4-quark operators, e.g.,
the $\Delta S=1$ lagrangian is given as
\begin{equation}
\label{74}
{\cal L}_{eff} =
\sum_i c_i (\mu ) {\cal O}_i ,
\end{equation}
where $c_i(\mu )$ are the Wilson coefficients
and ${\cal O}_i$ are local operators.
 As we have discussed in the preceding sections, there is
a complementary  representation in the form of
effective chiral lagrangian (\ref{61}). Both pictures
are equivalent. From the technical point of view
these are associated with definite problems:
using the lagrangian (\ref{74}), one cannot reliably calculate the
matrix elements of the composite local operators and
the use of the
complementary in ChPT gives rise to a large number of
undetermined couplings. The duality proposed in the
QCD hadronic duality sum-rule approach\cite{b56} spells
out the consistency of both representations; indeed,
there is a `window' where both pictures reliably
describe the same quantity. This is achieved
by writing down a system of finite-energy sum rules
(FESR's) which relate the integrals of the hadronic spectral
functions to the corresponding integrals calculated
in QCD.

The starting point is the two-point function
\begin{equation}
\label{75}
\psi (q^2) =
i \int d^4 x e^{i q\cdot x}
< 0 | T({\cal O} (x) {\cal O}^{\dagger} (0)) | 0 >,
\end{equation}
whose imaginary part enters the sum rules
\begin{eqnarray}
\label{76}
\Re_{n} (s_0) =
\int_0^{s_0} dt\;t^n\;\frac{1}{\pi}
Im\;\psi (t)_{hadronic}
= |g'(\mu )|^2 \int_0^{s_0} dt\;t^n \sum_\Gamma
| <0| \tilde{\cal L} |\Gamma > |^2 \nonumber \\
= \frac{s_0^{n+5}}{(n+5)(16\pi^2 )^3}
(\frac{\alpha_s (s_0)}{\alpha_s (\mu^2)})^\gamma
\{ a + a' \frac{\alpha_s}{\pi}(s_0) + b \frac{m_s^2 (s_0)}{s_0}
\nonumber \\
+ c \frac{m_s <0|\bar qq |0>}{s_0^2}
+ d \frac{<0|\frac{\alpha_s}{\pi}F_{\mu\nu}^a
F^{a\mu\nu} |0>}{s_0^2} + \dots \} ,
\end{eqnarray}
where $g' = c^{-1}(\mu )g$.
As can be seen from the l.h.s. of (\ref{76}),
the spectral function $\frac{1}{\pi} Im\psi (t)$
derived from the correlator (\ref{75}) describes,
 how
the operator ${\cal O}$ couples the vacuum
to physical states. In its ChPT version,
it enables one to calculate the threshold behavior
of the intermediate states ($\Gamma = K\pi ,\; K\pi\pi$ ,
etc.). The unknown constant $g'$ is factorized
out as an overall normalization. The upper limit
of integration, $s_0$, is the onset of the QCD
continuum. It should be high enough, so that the QCD
expansion on the r.h.s. of (\ref{76}) has sense.
The window we are talking about is roughly of the range
of $s_0$ where the two pictures are dual. Clearly,
this window cannot be of  large range;
for too low values of $s_0$, the QCD expansion
breaks, and for too high values, ChPT cannot reliably
parametrize the hadronic spectral function.

The parameters $a$ and $a'$ on the r.h.s. of (\ref{76}) are calculable
in perturbative QCD; $a$ gives the asymptotic
behavior in the chiral limit,
and $a'$ is a coefficient of  finite
$\alpha_s$ corrections. Mass corrections are
taken into account by the second term, and the third and  fourth
terms
are  nonperturbative corrections coming from quark and
gluon condensates, respectively. Leading logarithmic corrections
are summed up by making use of the renormalization group
equation, and they are factorized out as a ratio
of coupling constants to the power of $\gamma$, the latter
being proportional to the anomalous dimension of the
local operator. For the multiplicatively renormalizable
operators, $c^2(\mu^2)$ is of the form
\begin{equation}
c^2(\mu^2) = (\frac{\alpha_s(M_W^2)}{\alpha_s(\mu^2)})^{-\gamma},
\end{equation}
i.e., the $\mu$ dependence of the Wilson coefficient exactly
cancels the $\mu$ dependence of the matrix element.

In order to make sense, the power corrections in (\ref{76})
should not exceed $20-30\%$; otherwise,  higher mass
and/or condensate corrections become important.
This basically determines the allowed range
of $s_0$ from the point of view of perturbative QCD expansion.
It turns out that in the $K^0 - \bar K^0$ mixing
($B$ parameter)\cite{b56},
and in the $K^+\rightarrow \pi^+\pi^0$
decay ($\Delta I=3/2$
transition)\cite{b57}, this range is rather
high, $s_0 \sim 7-11\;GeV^2$. This implies that one has
to correct the pure ChPT behavior
of  hadronic spectral functions for the
formation of resonances. The proposal in \cite{b56} is
to modulate the final-state interaction through
the Breit-Wigner form and normalize it in such
a way that for $t=0$ it reduces to the
chiral limit value.

The onset of scaling is obtained by taking the ratio
of the two sum rules
\begin{equation}
r_n(s_0) \sim \frac{\Re_{n+1}(s_0)}{\Re_n(s_0)} ,
\end{equation}
so that the unknown constant $g'$ drops out.
The ratio $r$ is usually normalized in such a way
that that it equals $1$ for the asymptotic case
(no QCD). Then one looks for the duality region,
i.e., for the region where the function $r_n(s_0)$,
calculated in ChPT, agrees with the same function
calculated in QCD. Once the duality range of $s_0$ is fixed,
any of the sum rules $\Re_n$ leads to the value
of $g$, for any $s_0$ in the duality range. With $g$
determined, one can easily get the decay amplitude using the
ChPT lagrangian.

The $B$ parameter plays the role of the constant $g'$. It
has been calculated by Pich and de Rafael\cite{b56},
with the result
\begin{equation}
\label{79}
|B| = (0.33\pm 0.09) [\alpha_s (\mu^2)]^{2/9} .
\end{equation}
Compared with typical results obtained in the alternative
approach (large- $N_c$ expansion, lattice QCD)
the result (\ref{79}) appears smaller
roughly by a factor of  $2$, and, obviously, the
vacuum saturation larger by a factor of $3$.

The same method has been applied by Guberina, Pich and
de Rafael\cite{b57} to the $K^+\rightarrow \pi^+\pi^0$
decay and the result (\ref{79}) agrees
with experiment within a few percent.
 This should be considered as a
successful test of the duality approach. It should be noted
that especially this amplitude has almost always been
reproduced in different approaches within a factor
of $2$, but better accuracy is difficult to achieve.

Recently, Prades {\it et al.}\cite{b56} have improved
the calculation of $B$ by a careful analysis of the
hadronic parametrization. The result is somewhat
higher, $|B| = (0.39\pm 0.10) [\alpha_s (\mu^2)]^{2/9}$ .

Unfortunately, the success shown by the above results
does not pertain to the $\Delta I=1/2$ amplitude. The
calculation by Pich et al.\cite{b58} failed to
reproduce this amplitude by an order of magnitude.
It was quite difficult to understand this failure.
Recently, definite progress has been made\cite{b55}
which points out the reason for the failure.
Namely, in the calculations of the $B$ parameter
and the $\Delta I=3/2$ amplitude, finite $\alpha_s$
corrections were taken into account. They
were found to be moderate, and because of
technical complexity, they were not calculated
for the  $\Delta I=1/2$ amplitude, i.e.,
it was assumed that they were also  moderate
in this case. However, Pich has
recently shown\cite{b62} that this
assumption is premature. The perturbative finite
$\alpha_s$ corrections to the $\Delta I=1/2$ amplitude
are so huge that the whole perturbative expansion breaks.
This clearly shows that the problem is highly
{\bf nonperturbative}, and cannot be handled
in  a perturbative way.

The calculation performed in \cite{b62} includes two assumptions that
significantly reduce the complexity of calculation:
(i) The operators ${\cal O}_{\pm}$ in the bare weak lagrangian are
handled without mixing, i.e., as they are
multiplicatively renormalizable,
(ii) the penguin operator $Q_6$ is taken in the
large-$N_c$ limit in order to be multiplicatively
renormalizable. Then, the calculation shows
that for ${\cal O}_+$, finite $\alpha_s$ corrections
are moderate, and for ${\cal O}_-$ and $Q_6$, they
appear with the coefficients $47/5$ and $423/20$,
respectively. With the usual $\mu^2 = t$ rescaling
which eliminates all logarithms in the spectral function,
the corrections exceed $100\%$, even
\newpage
\noindent
at $t=10\;GeV^2$,
showing a clear breakdown of the
perturbation expansion\footnote{
It is interesting to note that this behavior persists
in in higher orders. In the large-$N_c$ limit,
the ${\cal O}(\alpha_s^4)$ corrections can be easily
computed, leading to the spectral function of the
penguin operator $Q_6$:
\begin{equation}
\psi (t)_{peng} \sim
[\alpha_s (t)]^{18/11}
\{ 1+24.28 \frac{\alpha_s (t)}{\pi}
+ 470.72 (\frac{\alpha_s (t)}{\pi})^2 + \cdots \} .
\end{equation}
}.

We would like to point out that the
discussed breakdown of the perturbative QCD
expansion for $\Delta S=1$ transitions
may have serious consequences. Clearly, it is
not possible to use the QCD duality approach
in calculating these transitions. However, both
the large-$N_c$ approach of Bardeen et al.
and the lattice calculation also start from the
Wilson expansion with the $\mu$ scale set at
hadronic level. The huge radiative corrections
{\bf both} in the Wilson coefficients and in the matrix
elements cast serious doubt on the validity
of the starting point in the above calculations.

\subsection
{Hadronic Matrix Elements on the Lattice}

Lattice QCD offers a method for calculating hadronic
matrix elements from first principles.
The technical difficulties  are however enormous.
Besides the usual problems caused by insufficient lattice size,
lattice spacing, and computer time, the matching of the lattice and
continuum operators is complicated by a conflict between the lattice
regularization and the chiral properties of the theory
\cite{mai}. Since the chiral symmetry is broken on the lattice
with Wilson fermions and the flavor content of the standard model
is broken on the lattice with staggered (Kogut-Susskind) fermions,
one must take a linear combination of lattice operators
in order to form the desired continuum operators. This
`mixing problem' is extensively discussed in ref.~\cite{ber}.

    The lattice calculation of the matrix element of a local
 operator between meson states is based on `measuring'
the corresponding three-point correlation
 function on the lattice:
\begin{equation}\label{cor}
G(x,0,y)=
<P_5(x){\cal O}(0)P_5(y)>.
\end{equation}
$P_5$ are diquark operators
with the same flavor content as the corresponding mesons,
and $\cal O$ is a 4-quark operator, e. g., appearing in
the effective lagrangian (\ref{74}).
Let us illustrate the
method on a typical matrix element, for  example
$<\pi^+|\bar{s}\Gamma u\bar{u}\Gamma d|K^+>$.
By replacing $\pi^+$ and $K^+$ by their diquark operators
we arrive at the correlation function
  \begin{equation}\label{cor2}
G(x,0,y)=
<\bar{d}(x)\gamma_5u(x)
\bar{s}(0)\Gamma u(0)\bar{u}(0)\Gamma d(0)
\bar{u}(y)\gamma_5s(y)>.
\end{equation}
The contraction of
the quark fields in (\ref{cor2})  in
all possible ways yields the `eight' and the `eye'
diagrams \cite{son} depicted in Fig. 2.
In  $K\rightarrow\pi\pi$, the eye
diagrams are pure $\Delta I = 1/2$, whereas the eight diagrams
may be either $\Delta I = 1/2$ or 3/2.
In the $\Delta S=2$ transition, only the eight
diagram contributes.
The propagators in Fig. 2 are quark propagators
 averaged over gauge configurations on the lattice.
The configurations which include internal fermion loops
are absent if the so-called `quenched' approximation
is used. Technically, the fermion determinant in
the averages is set equal to one. The approximation
is justified due to a $1/N_c$ suppression of the femion
loops. In the eye diagram, however, the $u$-quark loop
is not of higher order in $1/N_c$ and therefore the eye cannot be
eliminated on the basis of the quenched approximation.
Moreover, its contribution is believed to be responsible
for a large part of the $\Delta I=1/2$ enhancement.

As regards the $K\rightarrow\pi\pi$ decays,
there are basically two methods of calculation, depending
on the prescription used for lattice fermions.
In the first method, the $K\rightarrow\pi\pi$ amplitude is calculated
 directly, with all particles at rest and all
quarks degenerate.
All particles are on-shell, so the amplitude is well
defined but the weak Hamiltonian must insert momentum.
Lowest-order ChPT is then used to relate the
amplitude to the physical one. The method involves a relatively
small chiral extrapolation and  Wilson fermions can be
applied.
This method has been used by the Bernard and Sony group
 \cite{ber,b13,BS} and the European Lattice Collaboration (ELC)
\cite{mai,b12,ELC}.

The  second method is to measure the $K\rightarrow\pi$
and $K\rightarrow vac$ matrix elements and use ChPT
to relate these to the physical
 $K\rightarrow\pi\pi$ amplitude \cite{b54}.
In  ChPT, the physical amplitude is proportional
to the matrix element
$<\pi|{\cal O}_{sub}|K>$,
where the subtracted operator is defined as \cite{kil}
\begin{equation}\label{sub}
{\cal O}_{sub}={\cal O}
-\rho (m_s+m_d) \bar{s}d.
\end{equation}
The coefficient $\rho$ is determined from the ChPT relation
\begin{equation}\label{rho}
<0|{\cal O}|K>=
\rho (m_s-m_d) <0|\bar{s}\gamma_5d|K>.
\end{equation}
This method is appropriate for staggered fermions
owing to their good chiral properties, which
allow the use of equation (\ref{rho}).
In addition to calculating the eight and the eye diagrams, one
has to calculate the matrix elements of the
two-quark operator subtraction and the $K\rightarrow vac$
diagram needed to fix the coefficient of the
subtraction.
 This method has been mainly used by the staggered group
\cite{kil,sha,sha2}.

If one computes $K\rightarrow\pi$
or $K\rightarrow\pi\pi$ on the lattice, one has
to face up to dealing with the $\pi\pi$ interactions
in the final state (FSI). The FSI
may cause a significant enhancement of the
$\Delta I=1/2$ amplitudes, and suppression of the
$\Delta I=3/2$ amplitudes \cite{sha}. In addition,
there is a final state phase to be determined in
order to sum both  amplitudes to get the
physical $K\rightarrow\pi^+\pi^-$ rates.

In the last few years  calculations of the hadronic
matrix elements on the lattice have shown substantial
improvement. The systematic errors and finite size effects are
now under better control.
Let us quote some of the recent results most of which,
however, should
 still be considered as qualitative.

Relatively reliable data are obtained for the
$\bar{K}K$ amplitude. The ELC collaboration quotes two
values for the $B$ parameter at $a^{-1}=1.34$ GeV, depending on what
type of fit is used:
$0.91 \pm 0.11$ (linear fit)
and
$0.64 \pm 0.11$ (quadratic fit)
\cite{ELC}.
The staggered group observed a deviation from
the asymptotic scaling in the region of
$\beta=6/g^2$ from 6 to 6.4.
 They find that if the
lattice spacing is reduced by a factor of 2
(keeping the physical volume fixed),
$B$ drops from
$0.69 \pm 0.02$ to
$0.54 \pm 0.05$,
whereas the scaling requires a change of
about a few percent.
The extrapolation to $a=0$
with the anomalous dimension included
would give
$Bg^{-4/9}\approx 0.45 \cite{kil}$.
Bernard and Sony observed a similar
behavior.
Going from the lattice size
$16^3\times 40$ to
$24^3\times 40$ at $\beta=6$, they found that
$B$ dropped from
$0.86 \pm 0.24$ to
$0.69 \pm 0.12$ \cite{BS}.
A probable  explanation for this scaling violation
is that one still sees large
${\cal O}(a)$ effects.

As regards the
 $K\rightarrow \pi$ and
 $K\rightarrow \pi \pi$ amplitudes
 the data are less reliable.
 The ELC group used two methods of calculation.
The first one was to translate their $\bar{K}K$
 data into the $K \pi$ amplitude and the second was
 a direct $K\rightarrow \pi \pi$ calculation.
 The results
 for the $\Delta I=3/2$ amplitude
  obtained using the two methods are
 $A_{3/2}=(7.0 \pm 0.8)10^{-8}m_K$ and
 $(8.6 \pm 0.8)10^{-8}m_K$, respectively, to
 be compared with the experimental value
 $3.7 \cdot 10^{-8}m_K$.
 The ratio R obtained using the $K \rightarrow \pi$
 method is $12 \pm 5$ with 75 configurations, but
 unfortunately $12 \pm 12$ with 110 configurations.
  The direct method yielded a  worse result:
 $R=35 \pm 30$ \cite{ELC}.

Sharpe suggested that the
$\Delta I =1/2$ rule might be an
accumulation of factors of 1.5 - 2
due to different mechanisms \cite{sha}.
Taking into account an enhancement of $R$
arising from the FSI, the lattice calculation
should aim at about 62 \% of the experimental
value, i.~e., $R_{eye+eight}\approx 14$.
The staggered group finds that
the eight diagram alone gives
$R_{eight}=3.6 \pm 0.02$, but
according to their last data, the eye
contribution is unexpectedly consistent with zero
\cite{sha2}.


\newpage

\section
{Rare Kaon Decays}

\subsection
{$K_L\rightarrow \mu\bar\mu$ and $K_{L,S}\rightarrow
\gamma\gamma$ Decays}

As far as weak interactions are concerned, the
$K_L\rightarrow \mu\bar\mu$ proceeds partly via a box diagram
(diagram (a), Fig. 3), and partly via an $sdZ$ vertex (black
box in diagram (b), Fig. 3).
The first diagram is similar to the process $K_L\rightarrow
\gamma\gamma$, which proceeds via diagram (b), Fig. 3,
where $q$ denotes  up-type quarks, u, c, and t.
In fact, a major contribution to the decay rate for
$K_L\rightarrow \mu\bar\mu$ comes from the
higher-order electromagnetic process (digram
(d), Fig. 3), i.e., via a two-photon intermediate
state. This diagram actually dominates the imaginary
part of the amplitude ${\cal A}(K_L\rightarrow \mu\bar\mu )$,
relating it to the process $K_L\rightarrow \gamma\gamma$:
\begin{equation}
\label{81}
\Gamma_{K_L\rightarrow \mu\bar\mu}^{absorptive} \sim
\Gamma_{K_L\rightarrow \gamma\gamma}
\end{equation}
We would like to stress that eq.(\ref{81}) would, in principle,
provide us with a lower limit for the partial
width
\begin{equation}
\Gamma_{K_L\rightarrow \mu\bar\mu}\geq
\Gamma_{K_L\rightarrow \mu\bar\mu}^{absorptive},
\end{equation}
with $\Gamma_{K_L\rightarrow \mu\bar\mu}^{absorptive}$
being the `unitary bound'.

Although, naively, the processes $K_L\rightarrow \mu\bar\mu$
(diagram (a)) and $K_L\rightarrow \gamma\gamma$ (diagram (b))
would both have amplitudes of comparable strength,
${\cal A} \sim e^4 /M_W^2 \sim G_F \alpha$, experimentally,
they differ by a factor of $10^{-5}$, the reason being,
of course, the GIM suppression. The respective branching
ratios are given by\cite{b19}
\begin{eqnarray}
\label{83}
BR (K_L\rightarrow \mu\bar\mu ) =
(7.3\pm 0.4 \times 10^{-9}, \\
BR (K_L\rightarrow \gamma\gamma ) =
( 5.70 \pm 0.27 ) \times 10^{-4}.
\end{eqnarray}

Closely related to the $K_L\rightarrow \mu\bar\mu$
decay is the process $K^+ \rightarrow \pi^+ \nu\bar\nu$.
It receives contributions both from diagram (c) and diagram (e)
in Fig. 3. Especially, the unitary bound derived for
$K_L\rightarrow \mu\bar\mu$ would set constraints on the
$K^+ \rightarrow \pi^+ \nu\bar\nu$ decay, provided
the $K_L\rightarrow \mu\bar\mu$ decay rate were
dominated by the absorptive part of
$K_L\rightarrow \gamma\gamma \rightarrow \mu\bar\mu$
{\it and} the dispersive part were negligible\cite{b72}.

The effective lagrangian for the $d\bar s \rightarrow
\mu\bar\mu$ and $d\bar s \rightarrow \nu\bar\nu$
processes in zeroth order in strong interactions is
\cite{b71}
\begin{equation}
\label{85}
{\cal L} = \frac{G_F^2 M_W^2}{\pi^2}
\bar s_L \gamma_\mu d_L
( \tilde C \bar\mu_L \gamma^\mu \mu_L
- \sum_i \tilde D_i \bar\nu_L^i \gamma^\mu \nu_L^i ).
\end{equation}
The coefficients $\tilde C$ and $\tilde D_i$ (i = generation
index) are related, via the unitarity relation
$\sum_j V_{js}^* V_{jd} = 0$,
to the Inami-Lim coefficients $\bar C$ and
$\bar D_i$\cite{b71}
which are functions of the mass ratios
$x_j = m_{q_j}^2 / M_W^2$ and $y_i = m_{l_i}^2 / M_W^2$.


As we discussed earlier, there are potentially large long-distance
effects in the $K_L\rightarrow \mu\bar\mu$ decay.

It is usually assumed that the dispersive part of the amplitude
for $K_L\rightarrow \gamma\gamma \rightarrow
\mu\bar\mu$ is small compared with the absorptive
one\cite{b72}. If this were true, the limit
on the short-distance part of $K_L\rightarrow \mu\bar\mu$
would be obtained as
\begin{equation}
{\cal A} (K_L\rightarrow \mu\bar\mu )_{short-dist.}
\leq {\cal A} (K_L\rightarrow \mu\bar\mu )_{exp}
- {\cal A} (K_L\rightarrow \gamma\gamma \rightarrow \mu\bar\mu )_{absorp.}.
\end{equation}
However, if the dispersive part of ${\cal A} (K_L\rightarrow \gamma\gamma
\rightarrow \mu\bar\mu )$ were large, and with opposite sign
to ${\cal A} (K_L\rightarrow \mu\bar\mu )_{short-dist.}$,
the above constraint would no longer be valid.

The decay $K_L\rightarrow \mu\bar\mu$ has been studied
in ChPT\cite{b73}. The main contribution
comes from the diagram (a), Fig. 3, i.e.,
the decay proceeds via the $\gamma\gamma$ intermediate
state. The $P\gamma\gamma$ vertex is described by the
Wess-Zumino term. The corresponding integral
is logarithmically divergent and the theory
requires counterterms in order to renormalize
the divergences. The finite part of the counterterms
unfortunately cannot be determined, and one
relies only on the logarithmic terms.
This is certainly not quite reliable, although one
does not expect that this arbitrariness essentially
changes the conclusions.

If one performs renormalization in the $\overline{MS}$-scheme,
the two-photon intermediate-state contribution
is\cite{b73}
\begin{equation}
\Gamma_{K_L\rightarrow \mu\bar\mu} =
\Gamma_{K_L\rightarrow \gamma\gamma}
\frac{\alpha^2 \beta}{2\pi^2}
(\frac{m_\mu}{m_K} )^2 | {\cal A} |^2,
\end{equation}
where $Re\;{\cal A}$ contains the logarithmic term, which is,
of course, $\mu$ dependent\footnote{
If one were able to determine the finite part of the
counterterms, the $\mu$ dependence would disappear,
i.e., the counterterms are also $\mu$ dependent.} :
\begin{equation}
Re\;{\cal A} =
3 \ln (\frac{m_\mu^2}{\mu^2}) -7
+ \frac{1}{\beta} \{ \frac{1}{2}
\ln^2 (\frac{1+\beta}{1-\beta})
- \frac{1}{3} \pi^2 + 2 Li_2
(\frac{1-\beta}{1+\beta}) \} ,
\end{equation}
and $\beta = (1-4m_\mu^2 / m_K^2 )^{1/2}$.
On the other hand, $Im\;{\cal A}$ has no
$\mu$ dependence:
\begin{equation}
\label{89}
Im\;{\cal A} =
-\frac{\pi}{\beta} \ln (\frac{1+\beta}{1-\beta}).
\end{equation}
The amplitude ${\cal A}$ has been obtained by
using a once-subtracted dispersion relation with
$Im\;{\cal A}$ being known from the calculation performed
in ref.\cite{b74}.

Choosing $\mu^2 \approx m_K^2$, one gets
\begin{equation}
\frac{\Gamma_{K_L \rightarrow \mu\bar\mu}}
{\Gamma_{K_L\rightarrow \gamma\gamma}}
= 3.6 \times 10^{-5}.
\end{equation}
Theoretical result is about twice as large as the
experimental one. This, however, should be considered
as reasonable agreement, since:

(i) the calculation is not complete;
the finite part of the counterterms has not been
determined, and

(ii) the $\mu$ dependence of $Re\;{\cal A}$ is
pronounced.
In addition to the rather strong $\mu$
dependence (a factor of 4 in the range $0.2 \leq \mu \leq 1\;GeV$
), one
finds that, starting from $\mu \sim 0.3\;GeV$,
a dispersive part dominates over an
absorptive one, the former being a factor of 2 larger
than the latter, for $\mu^2 = m_K^2$.

Having in mind points (i) and (ii), it would be
premature to claim the real dominance of the
dispersive part. However, contrary to the usual assumptions, this
shows that it  may be significant.

Before we compare this calculation with the calculations
in different approaches, we should remember the following.
In principle, if the counterterms were determined, this would
be a complete calculation of
$\Gamma_{K_L \rightarrow \mu\bar\mu}$; i.e.,
it would also contain short-distance contributions.
In addition, all effects of heavy particles appearing in the theories
which go beyond the SM would also be included
in the counterterms.

{}~\\

{\it $K_S\rightarrow \gamma\gamma$ process in ChPT}
{}~\\
\addcontentsline{toc}{subsubsection}{$K_S\rightarrow
\gamma\gamma$ process in ChPT}\\

This process is described in ChPT by the lagrangians
(\ref{63})-(\ref{66}).
 In the calculation
performed in ref. \cite{b73}, only the
$\Delta I=1/2$ part in ${\cal L}_{\Delta S=1}$
has been kept. Then, the process proceeds via the diagrams shown
in Fig. 4.
It is  the simplest example of the predictive power
of chiral perturbation theory because it can arise only
from loop graphs involving ${\cal L}^{(2)}$, but does not receive
any contribution from the direct couplings $w_1$, $w_2$, $w_4$.
These couplings play the role of counterterms, i.e., the
loop divergences have to be absorbed in these couplings.
 Since couplings are absent, that means that
divergences must cancel and loops have to be finite.
The graphs displayed are expected to have both quadratic and
logarithmic divergences. The former are cancelled
by an $SU(3)$ invariant counterterm, and the latter
also disappear. This surprising result is a consequence of the requirement
that the amplitude vanishes in the $m_K^2 = m_\pi^2$ limit.
The final result is given as\cite{b73}
\begin{equation}
\label{91}
\Gamma_{K_S\rightarrow \gamma\gamma} =
\frac{\tilde g_8^2 \alpha^2 m_K^3 f_\pi^2}{2\pi^3}
(1 - m_\pi^2/m_K^2 )^2 |F(\frac{m_K^2}{m_\pi^2})|^2,
\end{equation}
where
\begin{eqnarray}
\label{92}
F(z) = \{ 1 - \frac{1}{z} [\pi^2 - \ln^2 Q(z)
- 2i\pi \ln Q(z)] \} , \nonumber \\
Q(z) = \frac{1 - (1-4/z)^{1/2}}{1 + (1-4/z)^{1/2}} .
\end{eqnarray}
The amplitude entering (\ref{91}) is dominated by the imaginary part,
 which is also
given in ref. \cite{b74}.
Numerically, eq.(\ref{91}) gives
\begin{equation}
\label{93}
\Gamma_{K_S\rightarrow \gamma\gamma} =
1.4\times 10^{-20}\;GeV.
\end{equation}
 From (\ref{93}) one infers that
\begin{equation}
\label{94}
BR (K_S\rightarrow \gamma\gamma )_{theory} =
2.0\times 10^{-6}.
\end{equation}
The measured branching ratio is\cite{b19}
\begin{equation}
BR (K_S\rightarrow \gamma\gamma )_{exp} =
(2.4\pm 1.2 )\times 10^{-6},
\end{equation}
in excellent agreement with the theoretical prediction (\ref{94}).
This is a nice example of how the full ChPT completely
accounts for the process in question

{}~\\
\newpage
{\it $K_L\rightarrow \gamma\gamma$ in ChPT}
{}~\\
\addcontentsline{toc}{subsubsection}{$K_L\rightarrow
\gamma\gamma$ in ChPT}\\

This decay basically proceeds via the $\pi^0$ , $\eta$, and $\eta '$
poles

\begin{equation}
\label{96}
{\cal A} = \sum_i \frac{{\cal A} (P_i\rightarrow \gamma\gamma )
< P_i | {\cal H} | K_L >}
{m_K^2 - m_\pi^2} .
\end{equation}
The $\eta '$ contributes only through an octet weak
transition. The vertices $P\gamma\gamma$ are again
described by the Wess-Zumino term.
The amplitude (\ref{96}) has been carefully calculated
in \cite{b73}, and by Goity \cite{b75}.
There are two sources of uncertainties that enter that
calculation. The first one is the $\eta -\eta '$
mixing angle $\theta$, whose previous value was
about $10^o$. However, recent measurements of $\eta$
production in $\gamma\gamma$ collisions\cite{b76}
give a higher $\eta\rightarrow \gamma\gamma$ rate, and
consequently a larger mixing angle $\theta$\footnote{
A large mixing angle, $\theta \approx 20^o$, is also
predicted in the large-$N_c$ limit\cite{b52}}.
The $\eta$ and $\eta '$ decays are
well described by taking the values\cite{b75}
\begin{equation}
\theta = 24^o \pm 2^o \;\;\;and\;\;\;
f_\pi /f_\eta ' = 1.04 \pm 0.05 .
\end{equation}
In spite of that, the result for
${\cal A} (K_L\rightarrow \gamma\gamma )$
is very sensitive to the parameters
$\delta = < K_L | \eta_8 >$ and
$\kappa = < K_L | \eta_1 >$.
Plotting the experimental $BR (K_L\rightarrow \gamma\gamma )$ in
terms of $\delta$ and $\kappa$ shows\cite{b75} that
for agreement with experiment, the parameter $\delta$
should deviate from its chiral limit value\footnote{
The good agreement with experiment, obtained for
$\theta = 13^o$ in ref. \cite{b73},
is, of course, spoiled with the present
value of $\theta$. The same is true for the
calculation of ref. \cite{b77}.}.

{}~\\

\subsection
{Radiative Kaon Decays in ChPT}

We have already discussed the decay $K_S\rightarrow
\gamma\gamma$, which is a nice example of the
self-consistency of ChPT. This decay belongs to
class (i) which includes  decays where the
contributions from  the counterterm lagrangian are forbidden
by some symmetry.
 Class (ii) includes decays where loops are finite
as in  class (i), but there is an additional contribution
of the counterterms that is now scale-independent.
 Finally, class (iii)  includes decays where
the loop amplitudes diverge. In this case, ChPT should allow
for the counterterms.

{}~\\
{\it Decays $K\rightarrow \pi\gamma\gamma$ }

\addcontentsline{toc}{subsubsection}{Decays $K\rightarrow \pi
\gamma\gamma$ }
{}~\\
The amplitude $K^0\rightarrow \pi^0\gamma\gamma$ can be
completely calculated in ChPT.
It is uniquely determined since, as in the process
$K_S\rightarrow \gamma\gamma$, the counterterms do not
contribute, and the amplitude is given in terms of the loop
amplitude. The spectrum is given by\cite{b53}
\begin{eqnarray}
\label{98}
\frac{d\Gamma (K_L\rightarrow \pi^0\gamma\gamma )}
{dy} =
\frac{4\alpha^2 m_K^5 \tilde g_8^2}
{(4\pi )^5}
\lambda^{1/2} (1,y,z^{-2})
| (y-z^{-2}) F(yz^2) + (1-y+z^{-2})F(y)|^2 \nonumber \\
y=m_{\gamma\gamma}^2 / m_K^2\;\;\;
[0\leq y \leq (1-z)^{-2} = 0.52  ] \nonumber \\
\lambda (a,b,c) =a^2+b^2+c^2 - 2(ab+bc+ca) ,
\end{eqnarray}
where $F(z)$ is the function given in (\ref{92}).
The spectrum  has a characteristic shape, quite different
from the phase space. The integrated rate (\ref{98}) gives
\begin{equation}
BR(K_L\rightarrow \pi^0\gamma\gamma ) =
6.8\times 10^{-7} .
\end{equation}
This branching ratio has recently been measured.
The reported result\cite{b19} is:
 $(2.0\pm 0.5)\times 10^{-6}$.

The parameter free prediction  is the ratio
\begin{equation}
\frac{\Gamma (K_L\rightarrow \pi^0\gamma\gamma )}
{\Gamma (K_S\rightarrow \gamma\gamma )}
=  5.9\times 10^{-4}.
\end{equation}

A similar process, $K^+\rightarrow \pi^+\gamma\gamma$,
belongs to  class (ii): contributions come
from a finite-loop amplitude, from  scale-independent
counterterms\footnote{
The counterterms enter the amplitude in the form
$\hat{c} = 32\pi^2 [4(L_9+L_{10}) -
\frac{1}{3} (w_1 +2 w_2 + 2 w_4 )]$, which is estimated
to be ${\cal O}(1)$\cite{b53}.},
and, in addition, from the anomaly.
The lower bound on the branching ratio is predicted to be
\begin{equation}
BR(K^+\rightarrow \pi^+\gamma\gamma ) \geq
4\times 10^{-7} ,
\end{equation}
and is expected to be measured soon, since the present
upper limit is\cite{b19}
\begin{equation}
BR(K^+\rightarrow \pi^+\gamma\gamma ) <
 10^{-6} .
\end{equation}

The decays $K\rightarrow \pi\gamma\gamma$ are also interesting because
 CP violation appears owing to the interference
between the absorptive
amplitude and the counterterms (complex numbers).
The charge asymmetry is given as
\begin{equation}\label{103}
\Gamma (K^+\rightarrow \pi^+\gamma\gamma ) -
\Gamma (K^-\rightarrow \pi^-\gamma\gamma ) =
Im\;\hat{c}\;1.5\times 10^{-23}\;GeV ,
\end{equation}
which crucially depends on the value of $\hat{c}$.
There is also a charge asymmetry for the
$K\rightarrow \pi e^+ e^-$ decays; it is given as
\begin{equation}\label{104}
\Gamma (K^+\rightarrow \pi^+ e^+ e^- ) -
\Gamma (K^-\rightarrow \pi^- e^+ e^- ) =
Im\;w_+\;1.6\times 10^{-25}\;GeV ,
\end{equation}
where $w_+$ is the following combination of counterterms
\begin{equation}
w_+ = -\frac{16\pi^2}{3}
(w_1 + 2 w_2 - 12 L_9 ).
\end{equation}

The source of  CP violation in the standard model is
presumably a phase in $g_8$. Turning on the electromagnetic
interaction induces the electromagnetic penguin as a
new source of CP violation. The latter contributes
only to the counterterm $w_1$ in the large-$N_c$
approximation. Since the counterterms $L_i$
are real,  the following relation holds:
\begin{equation}
Im\;\hat{c} = 2\;Im\;w_+ =
-\frac{32\pi^2}{3} Im\;w_1 .
\end{equation}
This relation predicts the ratio of charge asymmetry
(\ref{103}) versus charge asymmetry (\ref{104}) to be $\simeq 200$.

{}~\\
\newpage
{\it Decays $K_L\rightarrow \pi^0 e^+ e^- $ and
$\eta\rightarrow \pi^0\gamma\gamma$}
{}~\\
\addcontentsline{toc}{subsubsection}{Decays $K_L\rightarrow \pi^0
e^+ e^-$ and $\eta\rightarrow \pi^0
\gamma\gamma$ }\\

The one-photon exchange contribution to the amplitude
for $K_L\rightarrow \pi^0 e^+ e^-$ is purely
CP violating. The intrinsic CP violation is again
related to $Im\;w_1$ and is comparable with the `normal'
CP violation. However, what makes this amplitude
peculiar is the rate: the one-photon exchange
leads to $BR \sim 10^{-12} - 10^{-11}$.
Higher-order contributions in ChPT\cite{b53} are by
two orders of magnitude smaller than the one-photon
exchange
contribution\footnote{
As expected, correct introduction of vector mesons as nonlinear
realizations of chiral $SU(3)_V$ leads basically to the same
results as those obtained using counterterms.
The blind use of vector-meson dominance produces, e.g.,
contributions which are not permitted by
chiral symmetry\cite{b53}.}.

The electromagnetic suppression of the CP-conserving
amplitude makes it probable that this decay is dominated
by the CP-violating contributions coming from the small
CP-even $K^0_1$ component of the $K_L$ and through
the direct CP-violating contribution in
$K_2^0 \rightarrow \pi^0 e^+ e^-$. The latter
is of the same order or even larger then the indirect
one.

There is  another candidate, the $\eta\rightarrow
\pi^0\gamma\gamma$ decay, which causes difficulties. It receives
contributions
only from the finite loops, and the unique
prediction is
\begin{equation}
\Gamma (\eta\rightarrow \pi^0\gamma\gamma )_{one-loop}
= 0.35\times 10^{-2}\;eV .
\end{equation}
The experimental value is larger by two orders of
magnitude\cite{b19}:
\begin{equation}\label{108}
\Gamma (\eta\rightarrow \pi^0\gamma\gamma )_{exp}
= (0.85\pm 0.19)\;eV .
\end{equation}
Higher-order contributions are estimated,
and the result is still far  from the experimental
value. This decay really looks like an unexpected
failure of ChPT; however, before claiming it,
one should rather wait for confirmation of (\ref{108})
in independent measurements.

\newpage
\section
{Rare Decays $K^+ \rightarrow \pi^+ \nu\bar\nu$ and
$K^+ \rightarrow \pi^+ \tilde\gamma\tilde\gamma$}

\subsection
{Supergravity Effects in Rare Processes}

Given the fact that the limits on superparticle
masses are presently rather high,
roughly of the order of $M_W$
or higher,
it is clear that supersymmetric effects
are expected to be very tiny.
The strongest constraints seem to come
from the $\mu\rightarrow e\gamma$ process
in the lepton sector and the $K_L - K_S$
system in the quark sector.
The constraints imposed by these two processes
are in general respected in
$K_L\rightarrow \mu\mu$,
$K_L\rightarrow \mu e$,
$\mu\rightarrow e\bar ee$,
$\mu N\rightarrow eN$,
etc.\cite{b20,b21}

{}~\\

{\it Muon anomalous magnetic moment}
{}~\\
\addcontentsline{toc}{subsubsection}{Muon anomalous magnetic
moment}\\

No useful constraints are expected from the
$g - 2$ factor\cite{b44}.
The agreement between theory and experiment
for the anomalous magnetic moment of the muon is better than
$2\times 10^{-8}$. In the supersymmetrized self-energy
graph of the muon, the photino and the smuon enter the graph.
With $\mu_L$ and $\mu_R$ degenerate, the contribution
is proportional to the square of the muon mass.
Then, the effective vertex takes the form
\begin{equation}
\frac{e}{2m_\mu} F(q^2) \bar u \sigma_{\mu\nu}q^\nu u ,
\end{equation}
where $q$ is the momentum of the photon.
The $g - 2$ gets a contribution through
$F(0)$ which is proportional to the
muon mass. Besides, there are graphs
with wino and zino, Fig. 5.
A massless photino graph has been
calculated by Fayet\cite{b27}
with the result $a=1/2(g-2) \approx 10^{-9}$,
provided the smuon masses are larger than $15\;GeV$.
Adding wino and zino graphs, Ellis,
Hagelin and Nanopoulos\cite{b23}, and
Barbieri and Maiani\cite{b39} have calculated the full
contribution in global SUSY, with essentially the
same result. The extension to     a
more realistic supergravity model has
been made by Rom\~{a}o et al.\cite{b45}.
Contributions of neutral gauginos have terms
quadratic and linear
in $m_\mu$. The latter  are potentially large,
but
they are suppressed owing to a kind of GIM mechanism\cite{b23}.
At the end, the dominant
contribution comes from the photino graph,
which dominates by an order of magnitude.
For a given photino mass, $a$ is
a decreasing function of gravitino mass.
For $m_{3/2}\leq 15\;GeV$, the present agreement
between theory and experiment is preserved.
This means that supergravity effects
on $g - 2$  are really small.

{}~\\

{\it $\mu\rightarrow e\gamma $ decay}
{}~\\
\addcontentsline{toc}{subsubsection}{$\mu\rightarrow
e\gamma$ Decay}\\

In the standard model with massless neutrinos
the decay $\mu\rightarrow e\gamma$ is strictly
forbidden owing to lepton flavor conservation.
This also remains true in the supergravity model.
However, allowing for massive neutrinos compatible
with present experimental upper limits on their
masses, the branching ratio for this decay is
of the order of $\sim 10^{-16}$ in the standard model.
This is far below the experimental upper limit,
which is $7.2\times 10^{-11}$.

In the supergravity there is an additional diagram
for the $\mu\rightarrow e\gamma$ decay, diagram (a)
in Fig. 5.

Keeping the  same values of neutrino masses
and mixing angles as in the SM, and varying
gravitino and photino masses in the range
$20 - 250\;GeV$, one finds that for a given
$m_{3/2}$ the branching ratio increases
with $m_{\tilde\gamma}$, the reason being
that the increase of the photino mass results
in the decrease of the wino mass. Since
the branching ratio contains a term
proportional to $(M_W / M_{\tilde W})^2$,
it produces an enhancement for light
winos. This enhancement, however, weakens
with increasing gravitino
 mass\cite{b45}. Even for
a very light gravitino, $m_{3/2} = 20\;GeV$,
the SUSY branching ratio is smaller than
$10^{-13}$.

In view of the present experimental limits on
SUSY particles, these effects are also very tiny.

{}~\\

{\it $K_L - K_S$ System}
{}~\\
\addcontentsline{toc}{subsubsection}{$K_L - K_S$
System}\\

In the standard model, the $K_L - K_S$ system is described
by box graphs with $u$, $c$, $t$ quarks entering the
loops. For degenerate quarks, these
contributions cancel. The same remains true
for supersymmetrized graphs
since the partners of quarks and $W$ bosons,
squarks and gauginos,
have the same quantum numbers.

This cannot lead to the limit on squark masses,
but gives limits on mass differences
between the families. The bounds obtained
are even strenghtened if one
includes graphs with strong interacting gluinos.
The condition for this is that
superscalars have an off-diagonal
mass matrix in the basis where the quark mass
matrix is diagonal.
Thus, the bounds on masses  of
$\tilde d$, $\tilde s$, $\tilde b$ superscalars are obtained.
These bounds, coming from gluino exchange, are
generally by an order of magnitude
stronger than the corresponding ones
coming from simple supersymmetrized graphs.

The usefulness of bounds is often spoiled by
uncertainties in the QCD calculation of
matrix elements of the effective operators
once the heavy (super)fields are integrated
out. We discuss these problems in
Chapter 6.

\subsection
{The decay $K^+\rightarrow \pi^+ \nu\bar\nu$ in the
standard model}

At zeroth order in strong interactions
the decay $K^+\rightarrow \pi^+\nu\bar\nu$
proceeds through  diagrams in Fig. 6.
The first group of graphs, group (a), are box
graphs, which, calculated in the $R_\xi$-gauge, contain a
gauge-dependent part $\gamma (\xi )$.
The same function, $\gamma (\xi )$, also appears
in group (b) of diagrams, but with an opposite sign.
Therefore, the whole set of graphs in Fig. 6
is gauge independent, as it should be.
The whole set of graphs in Fig. 6 leads\cite{b70,b71,b79} to
the result (\ref{85}) for the effective lagrangian.

The branching ratio $BR (K^+ \rightarrow \pi^+ \nu\bar\nu )$
can be related to the $BR (K^+ \rightarrow \pi^0 e^+ \nu )$
by using an isospin relation\cite{b71}
\begin{equation}
BR (K^+ \rightarrow \pi^+ \nu_i\bar\nu_i )
= \frac{G_F^2 M_W^4}{4\pi^4}
\frac{|\tilde D_i |^2}{V_{us}^2}
BR (K^+ \rightarrow \pi^0 e^+ \nu_e ).
\end{equation}
Using the experimental values\cite{b19}
$BR (K^+ \rightarrow \pi^0 e^+ \nu_e )
= (4.82 \pm 0.06)\%$,$M_W = (80.22 \pm 0.26)$ GeV,
and $G_F = 1.16637 \times 10^{-5}$ GeV$^{-2}$,
one gets
\begin{equation}\label{111}
BR (K^+ \rightarrow \pi^+ \nu_i\bar\nu_i )
= 0.72 \times 10^{-6} \frac{| \tilde D_i |^2}
{V_{us}^2} .
\end{equation}
The function $\tilde D_i$ is proportional to the mass correction
function $\bar D_i$\cite{b71}, which is a function
of
mass ratios $x_j=m_{q_j}^2 / M_W^2$ and
$y_i= m_{l_i}^2 / M_W^2$.
Neglecting the
lepton mass in $y_i$, the value $\bar D_c$ for
$m_c = 1.5\; GeV$ is about
$4\times 10^{-3}$. This value decreases with increasing
lepton mass: for the $\tau$ lepton
with $m_{\tau} = 1.784\; GeV$,
$\bar D_c$ is reduced to $3.2\times 10^{-3}$.
However, the value of $\bar D_t$
increases almost linearly with $m_t$.
In the range $80\;GeV \leq m_t \leq 150\;GeV$,
$D_t$ is in the range
$1.38 \leq \bar D_t \leq 2.72$.
Taking a lower limit,
$m_t \geq 90\; GeV$, one gains
a factor of at least $400$
in the ratio
\begin{equation}
\frac{\bar D_t}{\bar D_c} \geq 400 .
\end{equation}

With neglect of the QCD corrections, (\ref{111}) becomes
\begin{equation}
BR (K^+ \rightarrow \pi^+ \nu_i\bar\nu_i )
= 0.72 \times 10^{-6} | V_{ud} |^2
| -\bar D_c (x_c) + \frac{V_{ts}^* V_{td}}
{V_{su}^* V_{sd}} \bar D_t (x_t) |^2,
\end{equation}
where the approximation $V_{cs}^* V_{cd} \approx -V_{us}^*
V_{ud}$ has been used. Neglecting lepton masses, one obtains
the function $\bar D(x)$ as
\begin{equation}
\bar D(x) = \frac{x}{4} - \frac{3}{4} \frac{x}{1-x}
+ \frac{1}{8} ( 1 + \frac{3}{(1-x)^2} -\frac{(4-x)^2}
{(1-x)^2} ) x\ln x .
\end{equation}

The weak contribution to the similar process
$K_L\rightarrow \mu\bar\mu$ is given by
\begin{equation}\label{115}
BR (K_L\rightarrow \mu\bar\mu )_{weak} \simeq
7.7\times 10^{-5} | V_{us} |^{-2}
| Re \sum_{i} V_{is}^* V_{id} \bar C_i (x_i) |^2,
\end{equation}
where $\bar C_i$ is defined by
\begin{equation}\label{116}
\bar C(x) = x + \frac{3}{4} \frac{x^2}{1-x} +
\frac{3x^2\ln x}{4(1-x)^2}.
\end{equation}
As we have discussed in  Chapter 4, the dominant contribution to
the $K_L\rightarrow \mu\bar\mu$ decay comes from the
two-photon intermediate state. Taking the absorbtive part
of the amplitude as calculated in  ChPT, eq.~(\ref{89}), and
{\bf assuming} no interference between the real part of the same
amplitude and the weak contribution as given in (\ref{115}),
one obtains
\begin{equation}
BR (K_L\rightarrow \mu\bar\mu )_{disp}
\simeq (2\pm 2)\times 10^{-9}.
\end{equation}
The above assumption then yields  the
constraint
\begin{equation}\label{118}
| Re\; V_{ts}^* V_{td} C_t (x_t) |^2
\leq 1.7\times 10^{-3}.
\end{equation}
The constraint in (\ref{118}) can be used keeping in mind
that the possible interference in the dispersive part
of the amplitude would weaken it.

\subsection
{QCD Corrections to the $K_L\rightarrow \mu\bar\mu$
and $K^+ \rightarrow \pi^+ \nu\bar\nu$ decays}

QCD corrections to the process $K^+ \rightarrow
\pi^+ \nu\bar\nu$
have been discussed in a number of
papers\cite{b29,b79}, \cite{b81}-\cite{b85}.
A controversy existed for some time,
and has been resolved
by Novikov et al. in \cite{b79}.

As far as the box diagram is concerned,
there are three types of gluon contributions
(diagrams (a) in Fig. 7):
(i) gluons might be exchanged
 between $d$ and $s$ quarks,
(ii) there is a quark-quark-W vertex correction, and
(iii) there is a self-energy graph
(mass renormalization) to an intermediate
quark.

It turns out that, working out QCD corrections
in the Landau gauge, only the self-energy graph
in Fig. 7 gives a logarithmic factor.

The loop integral
in the box diagram
 is of the form
\begin{equation}\label{121}
I = \frac{1}{2} m_c^2 M_W^4
\int \frac{dp^2}{(p^2 + M_W^2 )^2}
\frac{F(p^2)}{p^2 + m_c^2} \approx
\frac{1}{2} m_c^2
\int_{m_c^2}^{M_W^2}
\frac{dp^2}{p^2}F(p^2) .
\end{equation}
The last approximation
is valid only in the LLA.

An important point made by Novikov et al.\cite{b78}
is that the lower limit in (\ref{121})
is not  $\mu^2$ (arbitrary), but
$m_c^2$.
For $\mu^2 \leq m_c^2$, there are
no logarithmic gluon corrections
to the mass operator. This also
solves the controversy with
respect to previous calculations\cite{b81}-\cite{b83}.

As is obvious from the graphs (a) in Fig. 7, the
QCD corrections to the box graphs for
$K_L\rightarrow \mu\bar\mu$ and
$K^+ \rightarrow \pi^+ \nu\bar\nu$
are the same. This is not the case
for triangle graph corrections, to which
we turn next.

The triangle graph includes the $T$ product of three
hadronic currents. In contrast to the
box graph, there exist now two independent
distances. The calculation is
simplified because there exists
the Ward identity for the
$sdZ$ vertex. Therefore, one can reduce
the calculation of the matrix elements of
$K_L\rightarrow \mu\bar\mu$ to the
$T$ product of two currents and
the matrix element of the $T$ product of two currents and
the pseudoscalar
density $\bar c\gamma_5 c$.
Finally,   one gets
that the bare result for $K_L\rightarrow
\mu\bar\mu$ is multiplied by the
 Wilson coefficient
\begin{eqnarray}\label{126}
{\cal C}_{\mu\bar\mu}^c =
\frac{1}{2} \frac{4\pi}{\alpha_s (m_c^2)}
[\frac{3}{11} (\kappa_1^{12/25} - \kappa_1^{1/25})
+ 3(\kappa_1^{1/25} -1)
+ \frac{6}{7} (\kappa_1^{-6/25} - \kappa_1^{1/25})]
 \nonumber   \\
+ \frac{1}{2} (-\kappa_1^{12/25}
+2\kappa_1^{-6/25} + \kappa_1^{-24/25}),
\end{eqnarray}
where $\kappa_1 = \alpha_s (m_c^2) / \alpha_s (M_W^2)$.
The expression (\ref{126}) is given for  two
families. Its generalization to three
families is straightforward\cite{b29}.
The following comments are in order:
If $\alpha_s /4\pi \;\ln (M_W^2 / m_c^2 ) \sim 1$,
then the second term in (\ref{126}) is of the same order
as the neglected contributions in the first term.
In the SVZ calculation\cite{b78} the second term has been
kept for the following reasons:
(i) its extrapolation to the free-quark limit is smooth,
 (ii) the leading term is not so large numerically since
in the free-quark limit it vanishes, and (iii)
this remains true for the three families
as long as the $t$-quark is reasonably smaller
than $M_W$.

Generalization to three families
is straightforward.

The corresponding $c$-quark contribution is
given by ${\cal C}_{\nu\bar\nu}^c$
\begin{eqnarray}
{\cal C}_{\nu\bar\nu}^c =
\frac{1}{2} \frac{4\pi}{\alpha_s (m_c^2)}
[\frac{3}{11} (\kappa_1^{12/25} - \kappa_1^{1/25})
+ 12(\kappa_1^{1/25} -1)
 \nonumber  \\
+ \frac{6}{7} (\kappa_1^{-6/25} - \kappa_1^{1/25})]
+ \frac{1}{2} (-\kappa_1^{12/25}
+2\kappa_1^{-6/25} -3 \kappa_1^{-24/25}).
\end{eqnarray}

{}~\\

{\it QCD Corrections for $m_t \geq M_W$}
{}~\\
\addcontentsline{toc}{subsubsection}{QCD Corrections
for $m_t \geq M_W$}\\

The QCD corrections for the case $m_t \geq M_W$
are  difficult to handle, since each of
the subleading mass terms gets differently
`renormalized'. However, the leading term
for large $m_t$ is powerlike, i.e.,
the Wilson coefficients are simply
\begin{equation}
{\cal C}_{\mu\bar\mu} = {\cal C}_{\nu\bar\nu} =
\frac{1}{4},
\end{equation}
the dominant contribution coming from
the $sdZ$ graph. This is , however, true for very large
$m_t$; in practice, we use $m_t$ in the range
$80 - 200\;GeV$.
Therefore, for large $m_t$ it is better
to take the bare function $\bar D_t$ for
${\cal C}_{\nu\bar\nu}^t$ and similarly
for ${\cal C}_{\mu\bar\mu}^t$.

The final expression then becomes
\begin{equation}\label{131}
BR (K^+ \rightarrow \pi^+ \nu_i\bar\nu_i )
= 0.72 \times 10^{-6} | V_{ud} |^2
| -{\cal C}_{\nu\bar\nu}^c + \frac{V_{ts}^* V_{td}}
{V_{su}^* V_{sd}} {\cal C}_{\nu\bar\nu}^t |^2 .
\end{equation}

The quantity
$| V_{ts}^* V_{td} |^2$
is a very important factor entering the calculations
of rare decays. It appears as a `weight' to
the $t$-quark loop contributions. Using the
measured K-M matrix elements, one infers
from unitarity that $0.9985\leq | V_{tb} |
\leq 0.9993$ and
\begin{equation}\label{132}
|V_{ts} | \leq 0.054,\;\;\;
|V_{td} | \leq 0.024 ,
\end{equation}
which gives the constraint
\begin{equation}\label{133}
| V_{ts}^* V_{td} | < 0.0013 .
\end{equation}
It is important to note that any violation
of the relations (\ref{132}) would be a signal
of new physics beyond the standard model.
The derived constraint in (\ref{133}) will be
used to put an upper bound on
$BR (K^+ \rightarrow \pi^+ \nu\bar\nu )$.

In Fig. 8 we present the branching
ratio for the decay $K^+ \rightarrow
\pi^+ \nu\bar\nu$ as a function of
$m_t$. Since we have used the upper
limit (\ref{133}) on K-M matrix elements,
the result shown can be considered as
an upper limit. One sees that varying
$m_t$ in the range
$90\;GeV \leq m_t \leq 200\;GeV$
gives the branching ratio in the range
\begin{equation}\label{134}
BR (K^+ \rightarrow \pi^+ \nu\bar\nu )=
(1 - 4)\times 10^{-10}.
\end{equation}
The last Particle Data limit is\cite{b19}
\begin{equation}\label{519}
BR (K^+ \rightarrow \pi^+ \nu\bar\nu )<
3.4\times 10^{-8}.
\end{equation}

The following comments are in order:

(i) If one includes only the $c$-quark
contribution (without the QCD corrections),
one gets
$BR (K^+ \rightarrow \pi^+ \nu\bar\nu )= 1.1 \times 10^{-11}$,
for $m_c = 1.5\;GeV$.
Adding the QCD corrections lowers the branching ratio
to $3.1\times 10^{-12}$. Clearly, both numbers
are below the sensitivity of the BNL experiment E787
(a recent preliminary result is
$BR\leq 5\times 10^{-9}$, the expected
sensitivity is $2\times 10^{-10}$ ).

(ii) For larger values of $m_t$, the $t$-quark
contribution  dominates and the results
presented in Fig. 8 can be used in different
scenarios. If the experiment measures considerably
larger values than predicted, it is unlikely that
it could be explained by pushing $m_t$ too high,
since that would violate the
$K_L\rightarrow \mu\bar\mu$ bound (\ref{83}). However,
it would be possible to violate the unitarity
constraint in (\ref{133}) because, for example, of the
existence of the fourth family. Therefore,
any clear result beyond the predicted branching
ratio should be considered as a serious
signal of new physics. Clearly, further
improvements of the upper limits on $m_t$
and/or $| V_{ts}^* V_{td} |$ would further
constrain the range (\ref{134}).

(iii) In the case that $m_t$ is determined from
experiment, the eq. (\ref{131}) may be used to
test the standard-model prediction for the
branching ratio. On the other hand, the measured
branching ratio (\ref{83}) may be used to derive
an upper limit on  $m_t$.
Both possibilities are, of course, valid provided
no significant impact of `new physics' is
present.

(iv) We have kept  QCD corrections for
the $c$-quark contribution only. Since the
latter is not dominant, the errors entering
through the LLA, the charmed quark mass, etc.,
are not significant for the branching ratio.
For example, the threshold effects of quarks have
 recently been calculated\cite{b85} for
${\cal C}_{\nu\bar\nu}^c$ and they slightly modify
the resulting branching ratio.

(v) Recently, a penguin-box expansion has been performed
systematically\cite{b86}; cf. also \cite{b87}.
It enables one to perform a very
detailed analysis of $K\rightarrow \mu\bar\mu$ and
$K^+\rightarrow \pi^+ \nu\bar\nu$ with constraints coming from
the $\epsilon$, $\epsilon '$ analysis included, yielding
$BR(K^+\rightarrow \pi^+ \nu\bar\nu$ to be around $1\times 10^{-10}$.

\subsection
{$K^+ \rightarrow \pi^+ \nu\bar\nu$ decay
in the minimal supergravity model}

We  first supersymmetrize the box
diagram, i.e., the internal quarks and
W-bosons become squarks and gauginos
(Fig. 9a). This has to be compared with
the box diagram in the standard model
which is given as
\begin{equation}
{\cal A}_{SM}^{box} (
K^+\rightarrow \pi^+ \nu\bar\nu
) =-\frac{1}{M_W^2} \frac{g^4}{16\pi^2}
V_{ts}^* V_{td} [g(x_t,y_l) - g(x_u,y_l)]
(\bar s_L \gamma_\mu d_L )(\bar\nu \gamma^\mu \nu_L ),
\end{equation}
where $g(x,y)$ is given as
\begin{equation}
g(x,y) = \frac{1}{y-x} \{
-(\frac{x}{x-1})^2 \ln x +(\frac{y}{y-1})^2 \ln y
+ \frac{1}{x-1} - \frac{1}{y-1} \}.   \nonumber       \\
\end{equation}
The supersymmetrization leads to the following
expression:
\begin{equation}
{\cal A}_{susy}^{box}
(K^+\rightarrow \pi^+ \nu\bar\nu ) \simeq
\frac{1}{m_{\tilde W}^2}
\frac{g^4}{64\pi^2}
V_{ts}^* V_{td}
[g(\tilde x_t,\tilde y_l) - g(\tilde x_u,\tilde y_l)]
(\bar s_L \gamma_\mu d_L )(\bar\nu \gamma^\mu \nu_L ).
\end{equation}
Numerically, the ratio
\begin{equation}
\Re = | {\cal A}_{susy}^{box} / {\cal A}_{SM}^{box} |
\end{equation}
is of ${\cal O}(1)$. We have used $m_{\tilde t}^2 -
m_{\tilde u}^2 \simeq m_t^2$ and
$m_{\tilde W} = 40\;GeV$, $m_{\tilde l} = 40\;GeV$.
If one fixes $m_{\tilde u} = 70\;GeV$,
then $\Re$ grows with $m_t$;
for $m_t = 90\;GeV$, $\Re = 0.4$,
and for $m_t = 160\;GeV$, $\Re$ reaches 1.
For very large values of $m_t$, e.g., $m_t =200\;GeV$,
 $\Re = 1.4$.
With increasing $m_{\tilde u}$,
$\Re$ becomes somewhat smaller:
for $m_{\tilde u} = 100\;GeV$, $m_t =60\;GeV$,
$\Re =0.2$, and reaches 1 for $m_t = 200\;GeV$.

Although the SUSY  contribution may, in principle,
be a significant correction for large values
of $m_t$, it is unlikely, as noticed by Bertolini
and Masiero\cite{b47}, that this really would
happen. In fact, the large $\Re$ would mean the large SUSY
contribution to the $K_L\rightarrow \mu\bar\mu$ box.
This is  even more pronounced since the
Dirac algebra suppression in the $K_L\rightarrow \mu\bar\mu$ box
by factor of 4 does not appear in the SUSY version
and, morever, sneutrinos that enter the
$K_L\rightarrow \mu\bar\mu$ box further increase
the amplitude.

The promising contributions appear to be
spenguin graphs (b) and (c) of Fig. 9.
Graphs (b) with the external photon contribute,
for example, to the  $K^+ \rightarrow \pi^+
e^+ e^-$ decay. With the external $Z$-boson,
graphs (b) and (c) contribute to the
$K_L\rightarrow \mu\bar\mu$ and
$K^+ \rightarrow \pi^+ \nu\bar\nu$ decays.
For  set (b) it turns out that the self-energy
graphs exactly cancel with the vertex correction graphs. Therefore,
one has to include two L-R insertions in the squark
propagator\cite{b47}. The resulting mixing
$\tilde d_L - \tilde d_R$ is given by
$A m_{3/2} M_d$ (cf. (\ref{44})):
\begin{equation}
\left(
\begin{array}{cc}
m_{3/2}^2 + m_b^2 + cm_t^2  &  A m_{3/2} m_b \\
A m_{3/2} m_b  & m_{3/2}^2 + m_b^2
\end{array}
\right) .
\end{equation}

The standard-model value for
$K_L\rightarrow \mu\bar\mu$
is given by
\begin{equation}
{\cal A}_{SM} (K_L\rightarrow \mu\bar\mu )=
\frac{G_F}{\sqrt 2} \frac{\alpha}{\pi \sin^2 \theta_w}
V_{ts}^* V_{td} \; \bar C(x_t)
(\bar s_L \gamma_\mu d_L) (\bar\mu_L \gamma^\mu \mu_L),
\end{equation}
with $\bar C(x_t)$ given by (\ref{116}).
The SUSY contributes as\cite{b47}
\begin{equation}
{\cal A}_{susy}
(K_L\rightarrow \mu\bar\mu) \simeq
-\frac{G_F}{\sqrt 2} \frac{2\alpha_s}{3\pi}
V_{ts}^* V_{td} (A m_{3/2} m_b / m_{\tilde g}^2 )^2
[C(\tilde x_b) - (\frac{m_d}{m_b})^2
C (\tilde x_d)]
(\bar s_L \gamma d_L) (\bar\mu_L \gamma^\mu \mu_L),
\end{equation}
with $C(\tilde x)$ defined as
\begin{equation}
C(x) = \frac{1}{(x-1)^2}
\{ \frac{1}{2} - \frac{2}{x-1}
+ \frac{3}{(x-1)^2} \ln x - \frac{1}{x(x-1)} \}. \nonumber \\
\end{equation}
This leads to
\begin{equation}
\Re (K_L\rightarrow \mu\bar\mu ) \leq
\frac{2}{3} \frac{\alpha_s}{\alpha}
\sin^2 \theta_w (A m_{3/2} \frac{m_b}{m_{\tilde g}^2} )^2
| \frac{C(\tilde x_b)}{\bar C(x_t)} |.
\end{equation}
The corresponding expressions for
${\cal A} (K^+ \rightarrow \pi^+ \nu\bar\nu )$
can be obtained by changing $\bar C(x_t)$ to
$\bar D(x_t)$, i.e.
\begin{equation}\label{147}
\Re (K^+\rightarrow \pi^+ \nu\bar\nu ) =
| \frac{\bar C(x_t)}{\bar D(x_t)} | \Re (K_L\rightarrow
\mu\bar\mu ).
\end{equation}
Numerically, (\ref{147}) presents a sizeable correction only for small
values of the supergravity parameters.
For $m_{3/2} = 100\;GeV$, $m_{\tilde g} = 70\;GeV$,
$m_{\tilde b} = 40\;GeV$, and choosing $m_t = 90\;GeV$,
$\Re (K^+ \rightarrow \pi^+ \nu\bar\nu ) \simeq
0.3$. $\Re$ decreases with increasing $m_t$;
for $m_t = 150\;GeV$, $\Re = 0.14$.
On the other hand, larger values of $m_{\tilde b}$ suppress
$\Re$ significantly: for $m_{\tilde b} =  60\;GeV$,
$m_t = 60 GeV$, $\Re$ is lowered to 0.1.

Comparison with the supersymmetrized box shows
that for $m_t = 90\;GeV$, both supersymmetric
contributions are of the same order, $30-40 \%$.
With increasing $m_t$, $\Re^{spenguin}$ diminishes and
$\Re^{box}$ increases. Although a precise prediction
is difficult because of the arbitrariness of
SUSY parameters, it seems that in the mass range
$90\;GeV \leq m_t \leq 150\;GeV$
the SUSY effects\footnote{There is
a contribution of the graph where a nutral higgsino
plays the role of gluino; this contribution is, however,
suppressed by small Yukawa couplings.
This is not  the case for charged higgsinos:
although enhanced, this contribution is still smaller
than the contribution of the standard model\cite{b47}.}
may reach $50-100\%$, leading to the enhancement
in the branching ratio:
\begin{equation}
BR (SM\;+\;SUSY ) \sim (2-4) BR(SM).
\end{equation}

Prospects to detect the supersymmetry in this rare
decay are presently not very bright.
However, once $m_t$ and/or K-M matrix
elements are determined, the prediction (\ref{131}) for the
branching ratio in the SM is rather reliable.
The precise measurements may be able to
disentangle the possible departures from the
standard model. For example,
the branching ratio enhanced by a factor of 4 may be
due to supersymmetry, but also  to the fourth
generation; the precise determination, however, of the
K-M matrix elements plus unitarity could eliminate
the latter possibility.

\subsection
{$K^+ \rightarrow \pi^+ \tilde\gamma\tilde\gamma$ decay}

The $K^+$-decay into a pion and a pair of the
lightest superparticles is a very interesting
decay mode since it belongs to the `direct'-decay type:
the superparticles are produced as real particles.
However, in the minimal $N=1$ supergravity model
we are discussing, it is not likely that this mode will appear
for the simple reason: this decay is not allowed kinematically.
The underlying assumptions are:
that the $\tilde\gamma$ is the lightest
SUSY particle and $R$-parity is conserved.
The existing limits are the following\cite{b19}:
$m_{\tilde\gamma} > 15\;GeV$ for $m_{\tilde f} = 100\;GeV$,
$m_{\tilde\gamma} > 5\;GeV$ for $m_{\tilde e} = 55\;GeV$,
and there is no lower limit for $m_{\tilde e} > 58\;GeV$.
Although for the heavy selectron there is no lower
limit on $m_{\tilde\gamma}$, there is a cosmological
limit\cite{b19} that requires the photino to be heavier
than $300-500\;MeV$ or very light,
$m_{\tilde\gamma} \leq 100\;eV$.
Actually, there is a new allowed region\cite{b19}
$m_{\tilde\gamma} = 4-20\;MeV$ provided
the bound $m_{3/2} < 40\;TeV$ is satisfied.

The second underlying assumption (R-parity) may also be
relaxed. The SUSY models with $R$-breaking (which
allows the lightest superpartner to decay)
have been considered for a long time and
recently discussed again\cite{b47}.
Therefore, it looks meaningful to discuss the
possibility of the
$K^+ \rightarrow \pi^+ \tilde\gamma\tilde\gamma$
decay both from the experimental and theoretical
point of view.

The largest contribution to the decay comes
from the presence of large effects of
flavor-changing quark-squark-photino couplings, which
simply leads to a tree-graph $\bar s_L d_L
\rightarrow \tilde\gamma\tilde\gamma$ via
a $\tilde d_L^i$ exchange.
The result is very simple\cite{b47}
\begin{equation}\label{149}
{\cal A}
(K^+ \rightarrow \pi^+ \tilde\gamma\tilde\gamma) =
\frac{4\pi\alpha}{9}
[c m_t^2 / (m_{3/2}^4 + c m_t^2 m_{3/2}^2 )]
V_{ts}^* V_{td}
(\bar s_L \gamma_\mu d_L ) (\bar{\tilde\gamma} \gamma^\mu
\gamma_5 \tilde\gamma ),
\end{equation}
and has to be compared with the dominant
part of the
$K^+ \rightarrow \pi^+ \nu\bar\nu$
decay:
\begin{equation}\label{150}
{\cal A} (K^+ \rightarrow \pi^+ \nu\bar\nu ) =
-\frac{G_F}{\sqrt 2} \frac{\alpha}{\pi \sin^2 \theta_w}
V_{ts}^* V_{td} \bar D_t (x_t)
(\bar s_L \gamma_\mu d_L ) (\bar\nu_L \gamma^\mu \nu_L ).
\end{equation}
It is easily seen, that the former decay may well dominate over the
latter\footnote{
In order to compare both amplitudes, (\ref{150}) has to be
multiplied by $\sqrt 3$ for 3 neutrino species (the Majorana
nature of the photino is already taken into account
in (\ref{149}) ).}

Therefore, the experimental upper limit
on $K^+ \rightarrow \pi^+ +\; $`$nothing$' may be converted
to the lower limit on the squark mass which appears as
$m_{\tilde q}^{-2}$ in the expresion (\ref{149}).
Using the present experimental bound in (\ref{519})
one derives the following constraint on the
squark mass:
\begin{equation}
m_{\tilde q} > 50\;GeV .
\end{equation}
As we have said, this limit is valid provided, that the decay
$K^+ \rightarrow \pi^+ \tilde\gamma\tilde\gamma$
is allowed kinematically\footnote{
The tree-graph decay into  neutral higgsinos is
suppressed for the same reason as in the
spenguin contribution involving higgsinos.}.

\newpage
\section
{Constraints on Supergravity Parameters from $CP$ Violation
and $K\rightarrow \pi\pi$ Decays}

\subsection
{$K^0 - \bar K^0$
Mixing}

In this section we discuss how the
$K_L - K_S$ mass difference and the $\epsilon$
parameter in the $K^0 - \bar K^0$ system
constrain a general version of the minimal
$N=1$ supergravity model\cite{b42}.
The supersymmetric contributions to
$\Delta m_K$ and $\epsilon$ are
box graphs depicted in Figs. 10 and 11.
The graphs in Fig. 10 represent the mere
supersymmetrization of the `standard box':
the internal quarks are replaced by their scalar partners,
the squarks $\tilde u_i$, whereas the
$W$ bosons and the Higgs scalars are
substituted by their fermionic partners,
$\tilde W$ and $\tilde H$, respectively.
The gluon exchange diagrams in Fig. 11,
proportional to the strong coupling, are expected to
give the dominant contribution\cite{b35,b40}.
One usually assumes no great cancellation between
different contributions to $\Delta m_K$ and
$\epsilon$, and requires each contribution
to be less than the experimental values\footnote{
Franco and Mangano\cite{b34} gave the general
constraint assuming a coherent contribution of
 diagrams (a) and (b) in Fig. 11.}.

 Diagrams (a) in Fig. 10 yield a constraint\cite{b42}
\begin{equation}
(V^{\dagger} \frac{\Delta M_Q^2}{m_{3/2}}
V )_{12} \leq \frac{1}{100} \frac{m_{3/2}}{M_W} ,
\end{equation}
which implies near degeneracy amongst the
squarks. This is consistent with the usual
assumption about masses of  scalar fields
of  chiral superfields
\begin{equation}
M_{ab}^2 = m_{3/2}^2 \delta_{ab}
\end{equation}
at a renormalization point of the Planck mass.

The contributions of the diagrams (b) in Fig. 10
place a constraint on the flavor matrices
$\xi_U$, and  $\xi_D$
\begin{equation}
\xi_{U,D} = m_{3/2} A \lambda_{U,D} + \tilde\xi_{U,D}
\end{equation}
appearing in the trilinear scalar couplings (\ref{33}).
The $\tilde\xi_{U,D}$ is an arbitrary small matrix.
However, more stringent constraints on
$\tilde\xi_{U,D}$ are set by  diagrams (a) in Fig. 11,
leading to effective operators with mixed
L-R helicities, i.e., with the structure typical
of penguin graphs in the SM.
The local $\Delta S=2$ hamiltonian generated
by this graph is given by
\begin{equation}\label{155}
{\cal H}_{eff} =
\frac{1}{120} \alpha_s^2
\frac{M_{\bar SD}^2
M_{\bar DS}^{*2}}{m_{3/2}^6}
f (\frac{m_{\tilde g}^2}{m_{3/2}^2} )
(\theta_1 - 3\theta_2 ),
\end{equation}
where
\begin{equation}\label{156}
\left\{
\begin{array}{c}
\theta_1  \\  \theta_2
\end{array}
\right\}
= \bar s_L^i \gamma_\mu d_L^j\;
\bar s_R^k \gamma^\mu d_R^l \;
\left\{
\begin{array}{c}
\delta^{il} \delta^{jk} \\
\delta^{ij} \delta^{kl}
\end{array}
\right\}
\end{equation}
and
\begin{eqnarray}
f(x) = 20 \int_0^1 d\zeta\;\zeta^3 (1-\zeta )
[\zeta + x(1 -\zeta )]^{-3}, \nonumber \\
f(1) = 1.
\end{eqnarray}
The notation for the mass parameters in (\ref{155})
is explained in \cite{b42}.

In order to estimate the supergravity
contribution in $\Delta m_K$ and $\epsilon$,
we need an estimate of the matrix elements
of the operators $\theta_1$ and $\theta_2$.
Using the vacuum saturation as a starting
point,  we define \cite{b46}
\begin{equation}
< \bar K^0 | \theta_i | K^0 > =
\frac{1}{2} f_K^2 m_K^2
(1 + \frac{1}{N_c} ) B_{\theta_i} .
\end{equation}
With this definition, the vacuum saturation
values are $B_{\theta_1} = -9.6$ and
$B_{\theta_2} = -4$.
As usual, the L-R helicity structure
of our operators disqualifies the reliability
of the vacuum saturation estimate.
A better estimate is obtained by
making use of the QCD duality approach\cite{b56}.
We have to study the behavior of
the two-point function
\begin{equation}
\psi_{\theta_2}(q^2) =
i \int d^4 x e^{iq\cdot x}
< 0 | T(\theta_2 (x) \theta_2^{\dagger} (0)) | 0 > .
\end{equation}
The first step is to calculate
$\frac{1}{\pi}
Im\; \psi_{\theta_2} (t)$
both in QCD using the representation (\ref{156})
of $\theta_2$ and in ChPT using the
chiral representation
\begin{equation}
\theta_2^{chiral} =
B_{\theta_2} \frac{1}{3}
(\frac{f_K^2}{f_\pi^2})
\;:(i \frac{f_\pi^2}{2}
U\partial_\mu U^{\dagger})_{23}
\;(i \frac{f_\pi^2}{2}
U^{\dagger}\partial^\mu U)_{23}: .
\end{equation}

Next, we establish the duality region in terms
of FESR's. One needs two sum rules
to fix $s_0$, which is
the onset of the asymptotic QCD continuum.
These are
\begin{eqnarray}\label{161}
\Re_0 = \int_{4m_K^2}^{s_0}
dt \frac{1}{\pi} Im\;\psi_{\theta_2}(t), \nonumber \\
\Re_1 = \int_{4m_K^2}^{s_0}
dt \; t\frac{1}{\pi} Im\;\psi_{\theta_2}(t).
\end{eqnarray}
Once $s_0$ is fixed, any of the above sum rules
leads to the value of $B_{\theta_2}$.
The ratio of the sum rules
\begin{equation}
r = \frac{6}{5s_0}
\frac{\Re_1}{\Re_0}
\end{equation}
is a very sensitive test of duality.
It does not depend on $B_{\theta_2}$
and may be used to fix $s_0$.

Fig. 12 shows the results  plotted in
 both in ChPT and in QCD.
The departure from the asymptotic QCD
prediction (the dashed line) is due to mass
and condensate corrections. The dots are the values of $r$
in ChPT, and the solid line is the full
QCD result. The duality region is clearly
established in the range
$8 \leq s_0 \leq 11.5\;GeV^2$.
Any of the sum rules in (\ref{161})
leads to the value of $B_{\theta_2}$ once $s_0$
has been determined.

Fig. 13 shows the results obtained from the sum rule
$\Re_0$.
$B_{\theta_2}$ plotted as
a function of $s_0$ shows a plateau
behavior in the duality region.
The results obtained from the sum rule
$\Re_1$
are similar. The values for
$B_{\theta_2}$
 agree within $2\%$ in both cases.
Taking into account various theoretical
uncertainties\cite{b46},
we obtain the final estimate
\begin{equation}\label{163}
| B_{\theta_2} | = 0.25 \pm 0.15 .
\end{equation}

There are some principal difficulties
in determining the matrix element of the
operator $\theta = \theta_1 - 3 \theta_2$.
The QCD duality approach determines
only the absolute value of the $B$
parameter and a separate determination of
$B_{\theta_1}$
would not be very useful.
However, the operator $\theta_1$
is of higher order in ChPT with
respect to $\theta_2$
and is expected to be significantly smaller.
In addition, the QCD spectral functions of the
operators $\theta$ and $3\theta_2$
show very similar $t$-dependence
and are of about the same magnitude\cite{b46}.
Therefore, we may neglect the $\theta_1$
contribution
to a very good approximation.

Using our estimate (\ref{163}) of the matrix
element of the $\theta_2$ operator
and the experimental values for
$\Delta m_K$ and $\epsilon$,
we obtain the following constraints on
$M_{\bar SD}^2$
and
$M_{\bar DS}^{*2}$
:
\begin{eqnarray}\label{164}
Re\; (\frac{M_{\bar SD}^2 M_{\bar DS}^{*2}}
{m_{3/2}^4})
\leq 4.5\times 10^{-5}
\frac{m_{3/2}^2}{M_W^2}
[f(\frac{m_{\tilde g}^2}{m_{3/2}^2})]^{-1},  \\
Im\; (\frac{M_{\bar SD}^2 M_{\bar DS}^{*2}}
{m_{3/2}^4})
\leq 3\times 10^{-7}
\frac{m_{3/2}^2}{M_W^2}
[f(\frac{m_{\tilde g}^2}{m_{3/2}^2})]^{-1}  .\label{165}
\end{eqnarray}
These constraints are rather sensitive to the
variation of the ratio $m_{\tilde g} / m_{3/2}$.
If one takes $m_{\tilde g} = m_{3/2}$,
one finds that (\ref{164}) and (\ref{165})
are weaker by a factor of 5 than
the corresponding constraints of Dugan et al.\cite{b42}.

The superbox diagrams (b) in Fig. 11
yields the effective hamiltonian\cite{b35},\cite{b38}-\cite{b40}
\begin{equation}\label{166}
{\cal H}_{eff} =
\frac{\alpha_s^2}{36 m_{\tilde g}^2}
\sum_{i,j} S(x_i,x_j) V_{is}^* V_{id} V_{js}^* V_{jd}
(\bar s_L \gamma_\mu d_L)^2
\end{equation}
where the sum runs over indices $u$, $c$, and $t$,
and
\begin{equation}
x_1 =\frac{m_{\tilde d}^2}{m_{\tilde g}^2},\;\;\;
x_2 =\frac{m_{\tilde s}^2}{m_{\tilde g}^2},\;\;\;
x_3 =\frac{m_{\tilde b}^2}{m_{\tilde g}^2} .
\end{equation}
The down squark masses are related to each other
through the equation (\ref{47}).
Thus, one may assume that in a good approximation
$m_{\tilde d}^2 = m_{\tilde s}^2 =
m_{\tilde b}^2 + m_t^2$. The function $S(x,y)$
is given by
\begin{equation}
S(x,y) = 11\frac{K(x) - K(y)}{x-y}
+ 4\frac{I(x) - I(y)}{x-y}
\end{equation}
\begin{eqnarray}
K(x) = \frac{x^2 \ln x}{(1-x)^2} + \frac{1}{1-x} \nonumber \\
I(x) = \frac{x \ln x}{(1-x)^2} + \frac{1}{1-x}.
\end{eqnarray}
Using the unitarity of K-M matrix one finds the
super-box contribution to $\Delta m_K$
\begin{equation}\label{170}
(\Delta m_K )_{sbox} =
\frac{\alpha_s^2}{54m_{\tilde g}^2}
B
f_K^2 m_K | V_{ts}^* V_{td} |^2 \Delta S ,
\end{equation}
where
\begin{equation}
\Delta S = S(x_3,x_3) + S(x_1,x_1) - 2S(x_1,x_3),
\end{equation}
and $B=1$ for the vacuum saturation estimate of the LL operator
matrix element between the $K^0$ and $\bar K^0$ states.
For $m_t \ll m_{\tilde g}$, $\Delta S$
can be replaced by the second derivative of $S(x,y)$ and
(\ref{47}) simplifies, yielding an $m_t^4$ dependence of
$\Delta m_K$ and $\epsilon$ \cite{b34,b42}.
We shall, however, use the exact form (\ref{170}).

Eq.~(\ref{170}) can be used to place  lower bounds on the gluino
mass and down-squark masses, provided one has sufficient
information about the quark masses and K-M parameters.
Unfortunately, the top-quark mass $m_t$ is still not
known. The best one can do is to use the present
experimental lower bound
$m_t > 91\;$GeV\cite{b19} and the upper limit
$m_t \leq 180\;GeV$ from the standard model
constraints\cite{b88}. As regards the K-M parameters,
a rigorous upper bound given by (\ref{133}) can be used.
The lower bounds on $m_{\tilde g}$ and
$m_{\tilde d}$ for two values of $m_t$ are
as follows
\begin{center}
\begin{tabular}{ccc}
\makebox[2.3cm]{} &
\makebox[2.3cm]{} &
\makebox[2.3cm]{}     \\
$m_t$              &   100     &    150   \\
                   &           &          \\
$m_{\tilde g} >$   &   40      &    50    \\
                   &           &          \\
$m_{\tilde d} >$   &   105     &    160   \\
                   &           &          \\
\end{tabular}
\end{center}

The imaginary part of the strong superbox diagram (b)
in Fig. 11 yields a contribution to the $\epsilon$
parameter. Whereas for typical values
$m_{\tilde g} \simeq m_{\tilde d} \simeq
m_{3/2} = 100\;GeV$ $(\Delta m_K )_{sbox}$ is rather
small (of the order of $0.1 (\Delta m_K )_{exp}$ ), the
contribution to $\epsilon$ can be quite large\cite{b35}.
By making use of (\ref{166}), one finds that
\begin{equation}
|\epsilon_{sbox} | =
\frac{\sqrt 2}{108} \alpha_s^2
\frac{Bf_K^2 m_K}{m_{\tilde g}^2 \Delta m_K}
s_1^2 s_2^2 (s_2 s_3 s_\delta ) \Delta S ,
\end{equation}
which compared with $\epsilon_{exp} = 2.27\times 10^{-3}$,
and assuming $(\Delta m_K )_{sbox} / (\Delta m_K )_{exp}
\approx 0.1$,
leads to a stringent constraint on $s_2s_3s_\delta$:
\begin{equation}\label{174}
s_2s_3s_\delta \leq 2\times 10^{-5} \frac{1}{s_2^2} .
\end{equation}
This bound has important consequences for the contribution
of penguins and superpenguins to $\epsilon '$,
which we discuss in the next section.

\subsection
{CP Violation in the $\Delta S=1$ transition}

Supersymmetric contributions to $\epsilon '$ are
represented by the superpenguin operators depicted in Fig. 14.
Again, the corresponding diagrams involving winos are
neglected because they are proportional to $g^2$ rather than
to $g_s^2$. The superpenguin (a) in Fig. 14 leads to a phase in
the $K_L\rightarrow \pi\pi$ amplitude.
$\xi_{spen}$  compared with $\xi_{pen}$ from the standard
model equals\cite{b42}
\begin{equation}
\frac{\xi_{spen}}{\xi_{pen}} =
\frac{1}{5} (\frac{g_s}{g})^2
\frac{(m_t / m_{\tilde d} )^2}{\ln (m_t / m_c )^2} .
\end{equation}
This could be larger than 1 if $m_t$ is very large.
However, in this case, the superbox diagram dominates
$\epsilon$ and yields
a strong bound on $s_2s_3s_\delta$, (\ref{174}), which in its turn
makes the penguin contribution very small. Thus, the  conclusion
is that the superpenguin (a) in Fig. 12 may give a
significant contribution to $\epsilon '/\epsilon$ only in
a very small region of parameter space\cite{b42}.

The scalar mass insertions
$M_{\bar SD}^2$
and
$M_{\bar DS}^2$
which have been discussed in the preceding section,
also appear in the $\Delta S=1$
transition dipole moment operators\cite{b42},
represented by the superpenguin diagram (b) in Fig. 14.
The corresponding effective hamiltonian is
given by
\begin{equation}
{\cal H}_{eff} =
\frac{1}{32\pi}
\frac{\alpha_3 g_s}{m_{3/2}} {\cal F} (x)
[\frac{M_{\bar SD}^2}{m_{3/2}^2}
\bar s_R \sigma_{\mu\nu} F^{\mu\nu} d_L
+ \frac{M_{\bar DS}^{*2}}{m_{3/2}^2}
\bar s \sigma_{\mu\nu} F^{\mu\nu} d_R] ,
\end{equation}
with
\begin{eqnarray}
x = \frac{M_3^2}{m_{3/2}^2} \nonumber  \\
{\cal F} (x) =
\sqrt x [12 \Im (x) + \frac{4}{3} \frac{1}{x^2}
\Im (\frac{1}{x})] ,\;\;{\cal F}(1)=1+\frac{1}{9},   \nonumber  \\
\Im (x) = \int_0^1
d\zeta (1- \zeta )^2 [\zeta + x(1-\zeta )]^{-2} .
\end{eqnarray}
This effective hamiltonian could give a significant
contribution to $\epsilon '$ and to the
$\Delta I=1/2$ amplitude in the
$K\rightarrow \pi\pi$ decay.
Using, for example, the bag model estimate\footnote{
The bag model estimate may be rather crude for the
operators containing gluon fields.
The QCD duality approach is, however, very difficult to
apply here because of the huge $\alpha_s$ corrections,
typical of $\Delta I=1/2$ processes\cite{b62}.}
of the above transition
moment operators\cite{b89},
one finds the phase and the magnitude of the
$\Delta I=1/2$ amplitude
\begin{eqnarray}\label{178}
\xi_{trans\;mom} =
250
(\frac{100\;GeV}{m_{3/2}}) {\cal F}(x)
Im\;
(\frac{M_{\bar SD}^2}{m_{3/2}^2} + \frac{M_{\ DS}^{*2}}{m_{3/2}^2}), \\
a^{1/2}_{trans\;mom} =
1.2\times 10^{-5}
(\frac{100\;GeV}{m_{3/2}}) {\cal F}(x)
Re\;
(\frac{M_{\bar SD}^2}{m_{3/2}^2} + \frac{M_{\ DS}^{*2}}{m_{3/2}^2}).
\label{179}
\end{eqnarray}
The constraints (\ref{178}) and (\ref{179})
allow us to assume that the imaginary parts of
$M_{\bar SD}^2$ and $M_{\bar DS}^{*2}$
are much smaller than the real parts.
Assuming further that
$Re\;M_{\bar SD} \approx Re\;M_{\bar DS}$,
one obtains
\begin{eqnarray}\label{180}
| \epsilon '/\epsilon |_{trans\;mom} \leq
2\times 10^{-2} \Phi (x), \\
a^{1/2}_{trans\;mom} \leq
20\times 10^{-8} \Phi (x) , \label{181}
\end{eqnarray}
with
\begin{equation}
\Phi (x) =
 {\cal F}(x) [f(x)]^{-1/2}\;\;\;\Phi (1) = 1 + \frac{1}{9} .
\end{equation}
In view of the recently measured values
 $Re(\epsilon '/\epsilon ) =
(2.2 \pm 1.2)\times 10^{-3}$ \cite{b19},
the bound (\ref{180}) is not satisfactory.
On the other hand, the bound (\ref{181}), compared
with the experimental value
$a^{1/2} = 27\times 10^{-8}\;GeV$,
opens an interesting, although perhaps unrealistic,
speculation that a great deal of yet unexplained
$\Delta I = 1/2$ enhancement could be attributed
to the transition dipole effective operators
induced by the extended supergravity model.

The mass ratio $m_{\tilde g} / m_{3/2}$ is yet unknown.
Assuming $m_{\tilde g} = m_{3/2}$ and requiring
\begin{equation}
| \epsilon '/\epsilon |_{trans\;mom} =
14 \xi_{trans\;mom} \leq
3\times 10^{-3}.
\end{equation}
one finds the following constraint
\begin{equation}
Im\;
(\frac{M_{\bar SD}^2}{m_{3/2}^2} + \frac{M_{\ DS}^{*2}}{m_{3/2}^2})
\leq 8\times 10^{-7} (\frac{m_{3/2}}{100\;GeV}).
\end{equation}
This constraint is more stringent than the one obtained from (\ref{178}).

\newpage
\section
{Conclusions}

The processes discussed in
this review have been selected in such a way that they are
interesting from both the theoretical and experimental point of view.
The latter means that rare processes we are discussing
are expected to be detected in the near future.
Of particular theoretical interest are
 processes which either have significant corrections
(here `significant' means  corrections of the order ${\cal O}(1)$ )
owing to the supersymmetry, or the standard model predictions
are clear and unambiguous.  In the latter case, the departure
 may indicate physics beyond the standard model, and
in the former case
precise measurements  again check
the possible new physics.

The process $K^+\rightarrow \pi^+\nu\bar\nu$ receives
supersymmetric corrections which are of the order 1.
The resulting branching ratio is larger by a factor
of $2-4$ with respect to the standard model
prediction. The latter is a function of the K-M
parameters $V_{ts}^* V_{td}$
(only the upper limit is known),
and of the top-quark mass, for which we know
the lower limit. Partial knowledge of the above
parameters leads to the prediction
$BR(K^+\rightarrow \pi^+\nu\bar\nu ) =
(1 - 4)\times 10^{-10}$. The range given will
become more narrow if the lower limit on $m_t$ increases.
This decay is dominated by  short-distance dynamics, and
long-distance effects are expected to be small,
and its standard model prediction is relatively free of
usual uncertainties. Therefore,
if the decay is measured, it can  also be used as
a way to determine  $m_t$.

Direct production of superparticles (photinos) is
possible in the decay
$K^+\rightarrow \pi^+\tilde\gamma\tilde\gamma$,
provided it would not appear to be  kinematically
forbidden. The contribution is large enough, so that
the present upper bound on $K^+\rightarrow \pi +\;$`$nothing$'
leads to the constraint $m_{\tilde d} > 50\;GeV$.

Supersymmetric contributions to $\Delta m_K$ and $\epsilon$
may be significant for large values of $m_t$. This
yields constraints on gluino and squark masses as well
as the L-R helicity mixing mass parameters
$M_{\bar DS}$ and $M_{\bar SD}$. Further constraints
on these parameters follow from the requirement
that the contribution to $\epsilon '$ of the
transition dipole moment operators in the general
version of the supergravity model must not be too large.
In the constrained version of the model the superpenguin
contribution to $\epsilon '/\epsilon$ appears to
be rather small.

Given the fact that supersymmetric effects are rather
tiny, it is important to reduce uncertainties
in the calculation, especially the ones coming from
the nonperturbative (confinement) effects in QCD.
Some recent approaches, such as  chiral perturbation theory,
the large-$N_c$ expansion, QCD hadronic duality
sum rules, lattice QCD, are discussed. Particularly interesting
appear predictions of ChPT in rare decays.
Especially, the reachness of radiative $K$ decays clearly
shows  the predictive power of ChPT. Some
of these predictions are to be tested
in the forthcoming experiments in the near
future.

{}~\\

{\sc Acknowledgements}.
{\small One of us (B. G.) would like to acknowledge
the financial support by the
Bundesministerium f\"ur Forschung und
Technologie of the Fed. Rep. of Germany under
contract No. 0234 MU R, and to thank the Theory Group
at DESY in Hamburg for kind hospitality, where
the main part of this work was done.}

{}~\\
\newpage

\begin{small}

\end{small}
{}~\\

\newpage
\begin{small}
{\bf
Figure Captions
}\\

\addcontentsline{toc}{section}{
Figure Captions
}

{\bf Figure 1}. Typical graphs in ChPT. (I) Tree-level $K^0$
decay, (II) pion decay constant and wave function renormalization,
(III) loop corrections to the $K^0 - \bar K^0$ mixing,
(IV) loop corrections to the $K^+$ decay. The square is an insertion
of the weak lagrangian and the circle is a strong interaction
vertex.

{\bf Figure 2}. The `eight'(a) and the `eye'(b) diagram
contribution to the correlation between two mesons and a four
quark operator.

{\bf Figure 3}. Rare-decays graphs at the quark level,
(a) box graph for the $K_L\rightarrow \mu\bar\mu$ decay,
(b) box graph for the $K_L\rightarrow \gamma\gamma$ decay,
(c) $sdZ$-vertex graph for the $K_L\rightarrow \mu\bar\mu$ decay,
(d) two-photon contribution to the $K_L\rightarrow \mu\bar\mu$ decay,
(e) box graph for the $K^+\rightarrow \pi^+\nu\bar\nu$ decay.

{\bf Figure 4}. Rare decays in ChPT.
(a) two-photon contribution to the $K_L\rightarrow \mu\bar\mu$ decay,
(b) loop graphs for the $K_S\rightarrow \gamma\gamma$ decay,
(c) pole contribution to the $K_L\rightarrow \gamma\gamma$ decay.

{\bf Figure 5}. Supersymmetric contributions to
the $\mu\rightarrow e\gamma$ and $g-2$.
(a) $\mu\rightarrow e\gamma$ decay in SUSY.
(a), (b), and  (c): contribution to $g-2$.

{\bf Figure 6}. $K^+\rightarrow \pi^+\nu\bar\nu$ decay.
(a) box graphs, (b) $sdZ$-vertex contributions.

{\bf Figure 7}. QCD corrections to the
$K^+\rightarrow \pi^+\nu\bar\nu$ decay.
(a) corrections to  box graphs, (b) corrections
to the $sdZ$-vertex graph.

{\bf Figure 8}. Upper limit on
$BR (K^+\rightarrow \pi^+\nu\bar\nu )$
in the standard model as a
function of top-quark mass.

{\bf Figure 9}. Supersymmetric contributions to
the $K^+\rightarrow \pi^+\nu\bar\nu$ decay.
(a) supersymmetrized box graph.
(b) supersymmetrized $sd\gamma$ and $sdZ$ vertices.
(c) supersymmetrized $sdZ$ vertex with mass insertions.

{\bf Figure 10}. Supersymmetric box graphs proportional
to weak couplings.

{\bf Figure 11}. Supersymmetric box graphs proportional
to strong couplings.

{\bf Figure 12}. The ratio of the sum rules $r$
plotted versus $s_0$ in ChPT (dots) and in full
QCD (solid line).

{\bf Figure 13}. Results for  $B_{\theta_2}$ as a function
of $s_0$.

{\bf Figure 14}. Supersymmetric penguin graphs.

\end{small}

\end{document}